\documentclass[letterpaper]{article} 
\usepackage{aaai2026}  
\usepackage{times}  
\usepackage{helvet}  
\usepackage{courier}  
\usepackage[hyphens]{url}  
\usepackage{graphicx} 
\usepackage{natbib}  
\usepackage{caption} 
\usepackage[export]{adjustbox}
\usepackage{algorithm}
\usepackage{algorithmic}

\usepackage{svg}

\PassOptionsToPackage{dvipsnames,table}{xcolor}

\usepackage{booktabs,multirow,array}
\newcommand{\subhdr}[1]{\begin{tabular}{@{}c@{}}#1\end{tabular}} 

\usepackage{multirow}
\usepackage{graphicx}
\usepackage[dvipsnames, table]{xcolor}
\usepackage{colortbl}

\usepackage{makecell}
\usepackage{calc} 
\usepackage{svg}

\usepackage{pifont}
\newcommand{\cmark}{\ding{51}}
\newcommand{\xmark}{\ding{55}}
\usepackage{arydshln}
\usepackage{chngcntr}
\usepackage{amssymb}

\usepackage{listings}

\definecolor{dkgreen}{rgb}{0,0.6,0}
\definecolor{gray}{rgb}{0.5,0.5,0.5}
\definecolor{mauve}{rgb}{0.58,0,0.82}
\usepackage{mdframed}
\newmdenv[
  backgroundcolor=gray!10,  
  linecolor=black,          
  linewidth=1pt,            
  roundcorner=5pt,          
  nobreak=true              
]{custombox}

\usepackage{siunitx,xcolor,pgf}
\newcommand{\deltacol}[1]{%
  \begingroup
  \pgfmathparse{#1}%
  \ifdim\pgfmathresult pt>0pt \textcolor{teal}{\num{#1}}%
  \else\ifdim\pgfmathresult pt<0pt \textcolor{red}{\num{#1}}%
  \else \num{#1}\fi\fi
  \endgroup
}

\usepackage[inline,shortlabels]{enumitem}
\usepackage{siunitx}
\usepackage{booktabs, siunitx}
\lstdefinestyle{jsonprompt}{
  basicstyle=\ttfamily\small,
  showstringspaces=false,
  breaklines=true,
  breakatwhitespace=false,
  columns=fullflexible,
  keepspaces=true,
  postbreak=\mbox{\textcolor{gray}{\tiny$\hookrightarrow$}\space},
}

\title{
  \includegraphics[height=2em,valign=c]{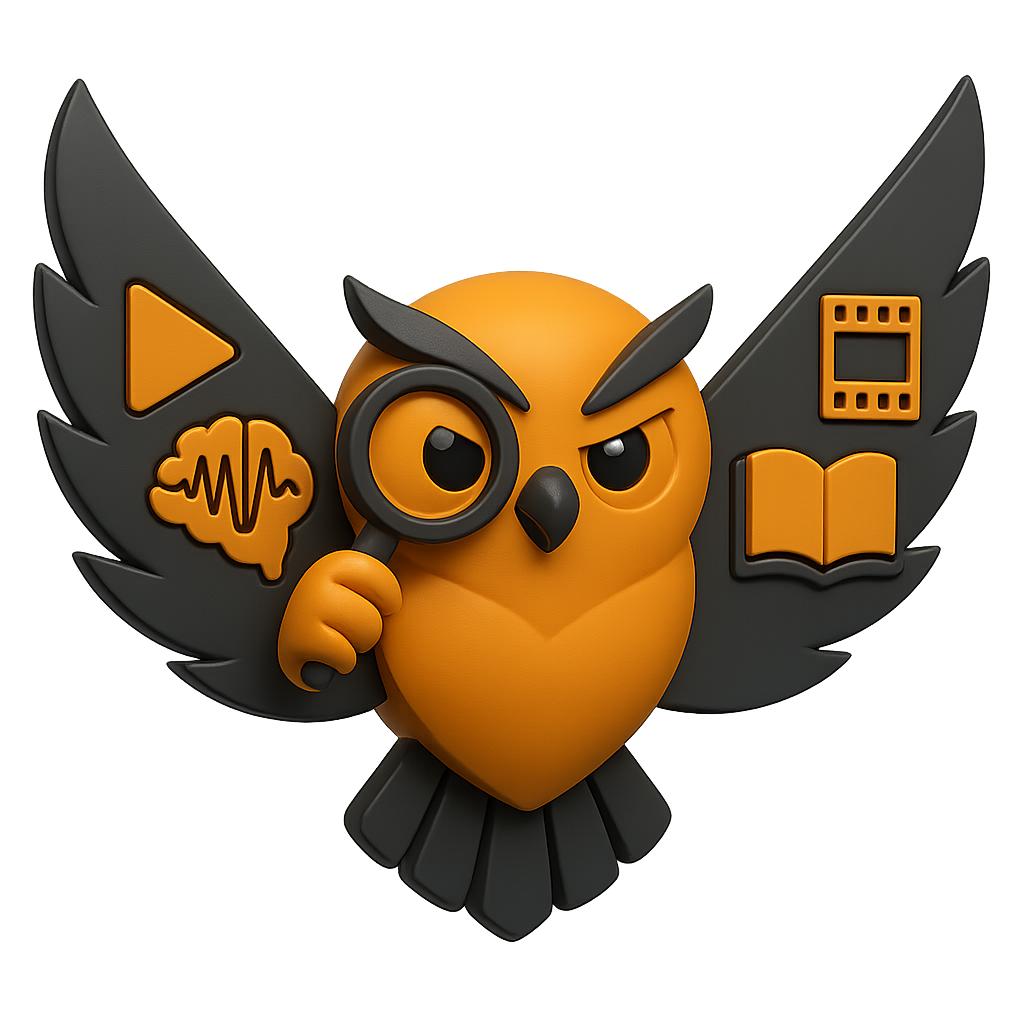} 
 MAVERIX: Multimodal Audio-Visual Evaluation and Recognition IndeX
}

\author{
    Liuyue Xie$^{*1}$ \quad
    Avik Kuthiala$^{*1}$ \quad
    George Z. Wei$^{*1}$ \\
    Ce Zheng$^1$ \quad
    Ananya Bal$^1$ \quad
    Mosam Dabhi$^1$ \quad
    Liting Wen$^1$ \\
    Taru Rustagi$^1$ \quad
    Ethan Lai$^1$ \quad
    Sushil Khyalia$^1$ \quad
    Rohan Choudhury$^1$ \\
    Morteza Ziyadi$^2$ \quad
    Xu Zhang$^2$ \quad
    Hao Yang$^2$ \quad
    László A. Jeni$^1$ \\
}
\affiliations{
    \textsuperscript{\rm 1}Carnegie Mellon University, \textsuperscript{\rm 2}Amazon, \textsuperscript{\rm *}Equal contribution\\
        
}

\usepackage{bibentry}
\usepackage{amssymb}
\usepackage{newunicodechar}
\newunicodechar{✓}{\checkmark}

\usepackage{newfloat}
\usepackage{listings}
\DeclareCaptionStyle{ruled}{labelfont=normalfont,labelsep=colon,strut=off} 
\floatstyle{ruled}
\newfloat{listing}{tb}{lst}{}
\floatname{listing}{Listing}

\usepackage{booktabs}

\begin{document}

\maketitle

\begin{abstract}
We introduce MAVERIX~(Multimodal audiovisual Evaluation and Recognition IndeX), a unified benchmark to probe the video understanding in multimodal LLMs, encompassing video, audio, text inputs with human performance baselines. 
Although recent advancements in models with vision and audio understanding capabilities have shown substantial progress, the field lacks a standardized evaluation framework to thoroughly assess their cross-modality comprehension performance. MAVERIX curates 2,556 questions from 700 videos, in the form of both multiple-choice and open-ended formats, explicitly designed to evaluate multimodal models through questions that necessitate tight integration of video and audio information, spanning a broad spectrum of agentic scenarios. MAVERIX uniquely provides models with audiovisual questions, closely mimicking the multimodal perceptual experiences available to humans during inference and decision-making processes. To our knowledge, MAVERIX is the first benchmark aimed explicitly at assessing comprehensive audiovisual integration in such granularity. Experiments with state-of-the-art models, including Qwen 2.5 Omni and Gemini 2.5 Flash-Lite, show performance around 64\% accuracy, while human experts reach near-ceiling performance of 92.8\%, exposing a substantial gap to human-level comprehension. With standardized evaluation protocols, a rigorously annotated pipeline, and a public toolkit, MAVERIX establishes a challenging testbed for advancing audiovisual multimodal intelligence.

\end{abstract}

\section{Introduction}
\label{sec:intro}

Human cognition seamlessly integrates visual and auditory information to reason, infer, and interact within dynamic environments. Replicating this ability in Multimodal Large Language Model (MLLM) systems remains a central challenge for AI, as autonomous agents must process complex audiovisual input to engage meaningfully with the world \cite{lin2023learning, llmsreason, testingnature}.

\begin{figure}[!t]
    \centering
     \includegraphics[width=\columnwidth]{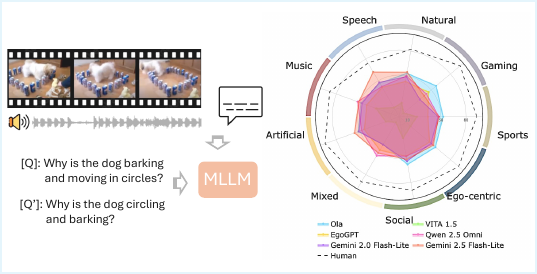}
     \caption{An illustration of our proposed benchmark, which includes highly audiovisual correlated questions and paraphrased questions, can be used to evaluate the model's underlying comprehension abilities and their gaps to humans. 
     }
     \label{fig:model_performance}
     \vspace{-10pt}
\end{figure}

\begin{figure*}[t]
    \centering
    \includegraphics[width=\textwidth]{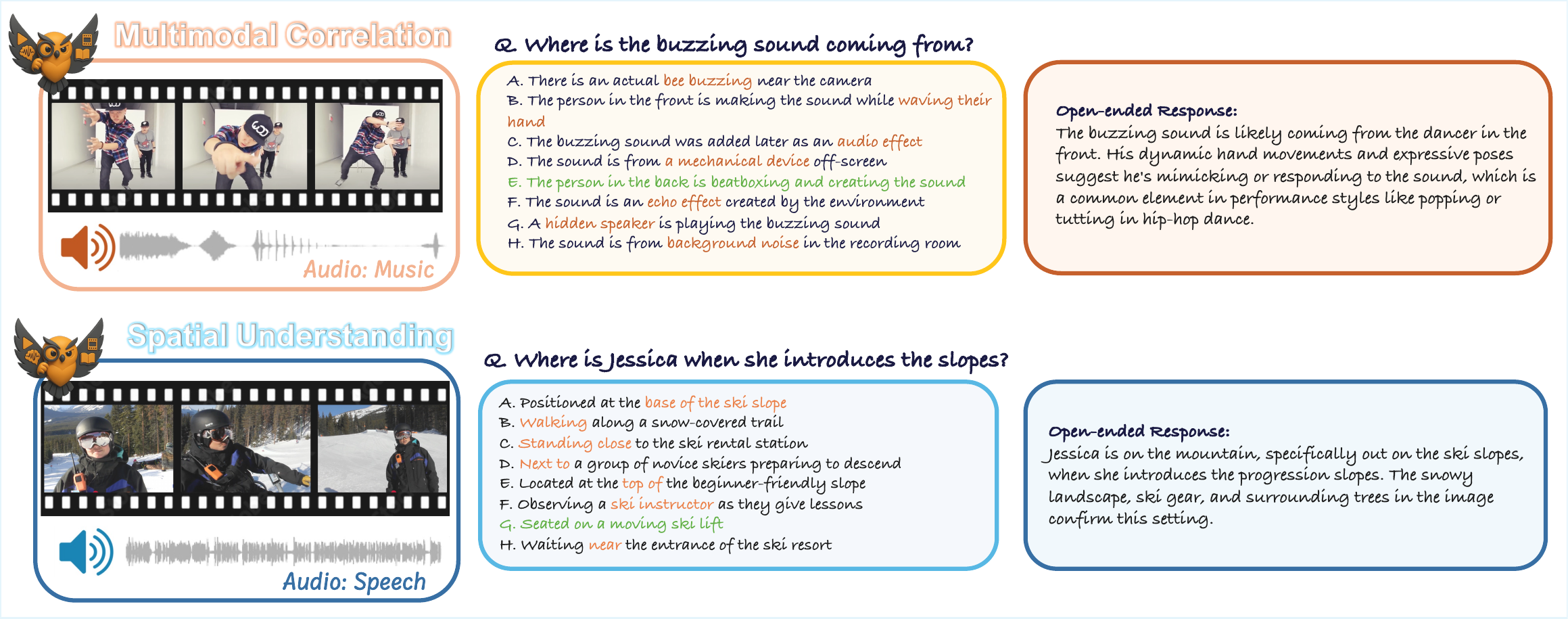}
    \vspace{-20pt}
    \caption{Example Agentic Categories and corresponding QAs in the MAVERIX benchmark.}
    \label{fig:example}
    \vspace{-10pt}
\end{figure*}

Recent progress in multimodal foundation models has brought us closer to this goal, but current benchmarks fall short in assessing their abilities to reason with multimodal inputs. Most benchmarks focus on static images \cite{chen2015microsoft, agrawal2019nocaps, li2025omnibenchfutureuniversalomnilanguage}, simple recognition, or questions that can be solved through unimodal cues, such as transcripts \cite{singh2019towards, chen2024far}. These benchmarks fail to probe the deeper, joint reasoning across modalities needed for real-world scenarios, such as interpreting social interactions or anticipating off-screen events \cite{hourvideo, mangalam2023egoschema}.

A core obstacle in designing effective multimodal benchmarks is ensuring that the questions genuinely require multimodal understanding rather than allowing models to exploit unimodal shortcuts or common sense from the training data. For benchmarks designed to expose the model understanding for highly multimodal data, their common adoption of a 4-way multiple-choice question for evaluation provides limited insights into the underlying interpretations \cite{li2025omnibenchfutureuniversalomnilanguage, hong2025worldsenseevaluatingrealworldomnimodal}. Many existing video-language benchmarks reduce to visible-object recognition or dialogue parsing, bypassing the need to synthesize audiovisual dependencies \cite{patraucean2023perception, kesen2023vilma, li2024mvbench}.

To address this, we introduce \textbf{MAVERIX}, a benchmark designed to evaluate multimodal video-audio understanding through questions that have tight modality interdependence. MAVERIX features questions from challenging agentic scenario categories: factual recall, causal understanding, sentiment analysis, temporal recall, situational awareness, context understanding, social interaction understanding, and emotional synthesis, covering 700 videos and 2,556 carefully designed questions. These are constructed through a hybrid human-AI pipeline to ensure that solving them requires intertwined audiovisual synthesis, revealing the underlying multimodal reasoning capabilities of models applied to the benchmark questions.

Evaluations of state-of-the-art proprietary and open-source models of different sizes, including Gemini 2.0/2.5 Flash-Lite(FL)\cite{geminiteam2024gemini15unlockingmultimodal}, GPT-4o \cite{openai2024gpt4ocard}, and Qwen 2.5 Omni \cite{xu2025qwen25omnitechnicalreport}, reveal significant gaps as shown in Fig.~\ref{fig:model_performance}. Gemini 2.5 Flash-Lite, even with direct audio-video inputs, achieves 54.7\% accuracy on multiple choice questions, significantly lower than human performance (92.8\%). Open-ended responses further expose weaknesses in temporal reasoning and contextual understanding, with models averaging only 1.9/5 vs. human 2.79/5 in GPT-4o-judged scoring. Further, models that are not capable of processing raw audio and rely solely on transcripts perform even worse, highlighting the inadequacy of text-only proxies for rich audiovisual comprehension \cite{ning2023video, fu2024video}.

By providing a unified evaluation framework, high-quality human-validated questions, and an open-source toolkit, MAVERIX aims to advance research toward robust multimodal reasoning at the human level.

\section{Benchmark Design and Construction}
\label{sec:benchmark}

MAVERIX challenges multimodal large language models (MLLMs) to \emph{integrate} audio and visual evidence under realistic conditions.
This section elaborates on four key aspects: (i)~the motivation behind our design, (ii)~the dataset construction pipeline, (iii)~the dual-format evaluation protocol, and (iv)~dataset statistics.
Fig. \ref{fig:workflow} visualizes the pipeline; Tab. \ref{tab:video_qa_benchmarks} compares the benchmark components with the relevant works; Tab. \ref{tab:subclass_stats} summarizes key dataset statistics.

\begin{table*}[t]
\centering
\scriptsize                    
\setlength{\tabcolsep}{6pt}    
\renewcommand{\arraystretch}{0.95}
\begin{adjustbox}{width=0.9\textwidth}  
\begin{tabular}{lcccccccccc}
\toprule
\textbf{Benchmark} & \textbf{\#Vid.} & \textbf{Med.\ Len.\,(s)} & \textbf{\#Q} &
\textbf{Mod.} & \textbf{MCQ.} & \textbf{\#Div.} & \textbf{Diff.} & \textbf{Shortcut} & \textbf{Human} & \textbf{OE.} \\
\midrule
MSRVTT-QA      & 2,990 &   15 &  72,821 & V       & 4-MCQ      & 1 & \textcolor{red}{\xmark} & \textcolor{red}{\xmark} & \textcolor{red}{\xmark} & \textcolor{red}{\xmark} \\
MSVD-QA        &   504 &    9 &  50,505 & V       & 4-MCQ      & 1 & \textcolor{red}{\xmark} & \textcolor{red}{\xmark} & \textcolor{red}{\xmark} & \textcolor{red}{\xmark} \\
ActivityNet-QA & 5,800 &   15 &  58,000 & V       & 4-MCQ      & 1 & \textcolor{red}{\xmark} & \textcolor{red}{\xmark} & \textcolor{red}{\xmark} & \textcolor{red}{\xmark} \\
How2QA         & 1,517 &   11 &  71,812 & V+S+A   & 4-MCQ      & 1 & \textcolor{red}{\xmark} & \textcolor{red}{\xmark} & \textcolor{red}{\xmark} & \textcolor{red}{\xmark} \\
\hdashline
AutoEval-Video &   327 &   32 &     450 & V       & 4-MCQ      & 1 & \textcolor{red}{\xmark} & \textcolor{red}{\xmark} & \textcolor{red}{\xmark} & \textcolor{teal}{\cmark} \\
TempCompass    &   410 &   11 &   1,540 & V       & 4-MCQ      & 2 & \textcolor{red}{\xmark} & \textcolor{red}{\xmark} & \textcolor{red}{\xmark} & \textcolor{teal}{\cmark} \\
Video-MME      &   900 & 1,072 &   2,700 & V+S     & 4-MCQ      & 3 & \textcolor{red}{\xmark} & \textcolor{red}{\xmark} & \textcolor{red}{\xmark} & \textcolor{red}{\xmark} \\
OmniBench      &   --    &   --    &    1142     & I+A   & 4-MCQ      & 3 & \textcolor{red}{\xmark} & \textcolor{teal}{\cmark} & \textcolor{red}{\xmark} & \textcolor{red}{\xmark} \\
WorldSense     &   1,662    &   141.1    &    3,172      & V+S+A   & 4-MCQ      & 3 & \textcolor{red}{\xmark} & \textcolor{red}{\xmark} & \textcolor{red}{\xmark} & \textcolor{red}{\xmark} \\
HourVideo      &   500 & 2,742 &  12,976 & V+S     & 4-MCQ      & 2 & \textcolor{red}{\xmark} & \textcolor{red}{\xmark} & \textcolor{red}{\xmark} & \textcolor{red}{\xmark} \\
\hdashline
\rowcolor{gray!11}
MAVERIX & 700 & 106 & 2,556 & \textbf{V+S+A} & \textbf{8-MCQ} & \textbf{7} & \textcolor{teal}{\cmark} & \textcolor{teal}{\cmark} & \textcolor{teal}{\cmark} & \textcolor{teal}{\cmark} \\
\rowcolor{gray!11}
MAVERIX-Long   & 700 & 345 & 2,556 & \textbf{V+S+A} & \textbf{8-MCQ} & \textbf{7} & \textcolor{teal}{\cmark} & \textcolor{teal}{\cmark} & \textcolor{teal}{\cmark} & \textcolor{teal}{\cmark} \\
\bottomrule
\end{tabular}
\end{adjustbox}
\caption{\textbf{Comparison with prior video-question benchmarks.}  
\textbf{Mod.}: V (video), S (subtitles), A (audio).  
\textbf{Ans.}: 4-MCQ, “8-MCQ+OE’’ (eight-option plus open-ended).  
\textbf{Diff.}: crowdsourced or expert difficulty labels. \textbf{\#Div.}: Number of division types. 
\textbf{Shortcut}: dataset validated against audio-only / video-only ablations.  
\textbf{Human}: ✓ if a benchmark reports any human baseline.}
\label{tab:video_qa_benchmarks}
\end{table*}

\subsection{Design Motivation and Principle}
\label{sec:motivation}

While previous video-understanding benchmarks curate multiple-choice question-answer pairs over different topics, some still suffer unimodal shortcuts and are limited in exposing models' underlying biases. For evaluating the models' multimodal understanding abilities, we source videos that capture a wide range of temporal events, spatial motions, and audiovisual correlations. While sourcing the videos and constructing the multiple choice and open-ended question-answer pairs, we follow the following design principles. 

\noindent\textbf{Avoid Unimodal Shortcut.}  
Many existing image and video based question answering benchmarks (e.g., TVQA, MSR‑VTT Q\&A) contain questions that can be answered from captions or a single salient frame, enabling unimodal shortcuts.

\noindent\textbf{Wide range of evaluation dimensions.}  
Deploying MLLMs in the real world requires that the models understand and handle scenarios with different skills. Single‑skill benchmarks do not reflect the breadth of reasoning required in open‑world settings. The curated benchmark measures the models in six evaluation dimensions, covering acoustic understanding, agentic skills, understanding by broad and sub-topic taxonomy, temporal understanding, and multimodal synthesis abilities. 

\noindent\textbf{Prevention of guess inflation.}  
We design a hybrid of eight-way multiple-choice, and open-ended QA for the benchmark, such that by-passing the questions with model-inherent biases can be evidently exposed. The hybrid design evaluates the models' actual abilities to interpret the input sources in different modality settings and provides a fair evaluation of their capabilities.

\subsection{Dataset Generation Pipeline}\label{sec:pipeline}



\noindent\textbf{Video Collection.}
We primarily source our video content from five datasets: YouTube‑8M\cite{haija2016youtube8m}, MSR‑VTT\cite{7780940}, UR‑FUNNY‑V2\cite{hasan-etal-2019-ur}, Ego4D\cite{ego4d}, and AudioSet\cite{jort_audioset_2017}. YouTube‑8M\cite{haija2016youtube8m} and AudioSet\cite{jort_audioset_2017} are large-scale datasets covering a wide range of taxonomies, where the videos have strong audiovisual correlations. MSR-VTT comprises of high quality video descriptions designed for video translation QAs. UR-FUNNY-V2 exhibits videos of different emotional states, challenging the models in their sentimental understanding. Lastly, Ego4D consists of egocentric, long-duration videos to probe model's understanding to daily interactions for agentic scenarios. The videos are selected by human annotators according to the principles described above, ensuring a thorough distribution across the topics, with different durations: \emph{short} with $<1$ min, \emph{medium} with 1-10 mins, and \emph{long} with 10-65 mins. Each video is processed and accompanied by subtitles generated with Whisper-v3~\cite{radford2022whisper} to ensure a fair evaluation on video-text models without audio-support.

\noindent\textbf{Initial Question Answering Annotation.}
A team of 8 expert annotators engaged in the initial question-answering pair curation. The annotators provided at least one question answer pair to each video to generate the initial ground truths. Then the same pair is expanded into eight-way multiple choice question with alternative distractive answers. 

\noindent\textbf{Shortcut Removal and Validation.}
Following the initial annotation, we use a semi-automated approach to validate difficulties with MLLMs and refine the questions to avoid any potential shortcuts. Each question undergoes three ablation tests with GPT‑4o-mini and Gemini 2.0-FL: \emph{text‑only}, \emph{video‑only}, and \emph{videos+subtitles}.  
If any ablation yields the correct answer for both models, the item is flagged and revised to reduce reliance on unimodal cues. 
All revisions are logged, and the final set is approved after a second expert pass. For example, a valid question might ask, \textit{``Why did the mechanic abruptly stop speaking?"} requiring both visual cues (e.g., discovering a leak) and audio cues (e.g., sudden silence). This protocol ensures MAVERIX's QA pairs demand genuine modality interdependence, preventing reliance on any single modality.

The difficulty labels were crowd-sourced through Amazon MTurk service~\cite{mturk} with 219 unique participants for gauging common consensus, and are determined based on the subtlety of cross-modal cues, the depth of understanding required, and the ease of locating relevant information in the video. The human performance evaluations were gathered through MTurk with 382 participants answering a 1/3 subset of the MCQs and open-ended questions.


\begin{table*}[t]
\centering
\setlength{\tabcolsep}{6pt}
\renewcommand{\arraystretch}{1.01}
\resizebox{0.9\textwidth}{!}{%
\begin{tabular}{l|ccc|ccc|cc|c}
\toprule
\multicolumn{1}{c|}{\textbf{Statistic}} &
\multicolumn{3}{c|}{\textbf{Audio Type}} &
\multicolumn{3}{c|}{\textbf{Agentic Categories}} &
\multicolumn{2}{c|}{\textbf{Topic Domain}} &
\multicolumn{1}{|c}{\textbf{Overall}} \\
\midrule
\multicolumn{1}{c|}{\textbf{Sub-class of QA}} &
\multicolumn{1}{c}{\subhdr{Mixed\\ Sound}} &
\multicolumn{1}{c}{\subhdr{Speech}} &
\multicolumn{1}{c|}{\subhdr{Artificial\\ Sound}} &
\multicolumn{1}{c}{\subhdr{Information\\ Querying}} &
\multicolumn{1}{c}{\subhdr{Egocentric\\ Agent}} &
\multicolumn{1}{c|}{\subhdr{Sentiment\\ Analysis}} &
\multicolumn{1}{c}{\subhdr{Humanity \&\\ Society}} &
\multicolumn{1}{c|}{\subhdr{Business \&\\ Commerce}} &
\multicolumn{1}{|c}{ } \\
\midrule\midrule
\multicolumn{1}{c|}{\textit{Agentic Abilities}} &
\multicolumn{3}{c|}{ } &
\multicolumn{3}{c|}{ } &
\multicolumn{2}{c|}{ } &
\multicolumn{1}{|c}{ } \\
\midrule
Causal Relationship        & 66  & 111 & 15 & 75  & 27  & 33 & 27 & 21 & 201 \\
Emotional Inference        & 51  & 57  & 18 & 66  & 21  & 27 & 27 & 12 & 129 \\
Factual Recall             & 516 & 690 & 60 & 672 & 171 & 186& 210& 120& 1311 \\
Situational Understanding  & 27  & 27  & 10 & 33  & 9   & 12 & 12 & 6  & 70 \\
Context Understanding      & 309 & 414 & 45 & 237 & 147 & 138& 111& 84 & 771 \\
\midrule\midrule
\multicolumn{1}{c|}{\textit{QA Lengths}} &
\multicolumn{3}{c|}{ } &
\multicolumn{3}{c|}{ } &
\multicolumn{2}{c|}{ } &
\multicolumn{1}{|c}{ } \\
\midrule
Question                   & 11.26 & 11.63 & 9.06 & 10.31 & 10.41 & 10.00 & 10.09 & 10.51 & 11.28 \\
Options                    & 11.76 & 10.52 & 11.99& 9.92  & 12.29 & 12.89 & 10.30 & 12.54 & 11.13 \\
Open-ended answer          & 13.30 & 12.69 & 11.68& 10.46 & 12.77 & 13.51 & 12.73 & 12.12 & 12.85 \\
Subtitle length            & 440.79& 682.24& 419.87& 485.08& 1257.19& 360.40& 351.32& 488.34& 558.06 \\
\midrule\midrule
\multicolumn{1}{c|}{\textit{Video–Audio Statistics}} &
\multicolumn{3}{c|}{ } &
\multicolumn{3}{c|}{ } &
\multicolumn{2}{c|}{ } &
\multicolumn{1}{|c}{ } \\
\midrule
Media Length               & 319.98 & 381.24 & 327.25 & 315.77 & 1039.18 & 289.09 & 222.42 & 259.69 & 352.63 \\
Media min. Length          & 6.15   & 10.03  & 5.57   & 5.57   & 10.04   & 5.57   & 5.57   & 10.15  & 5.57 \\
Media max. Length          & 6620.63& 4427.76& 3205.50& 6620.63& 6620.63 & 3851.93& 1800.17& 511.23 & 6620.63 \\
Media std.                 & 527.50 & 681.63 & 482.65 & 603.49 & 1341.37 & 439.54 & 268.20 & 117.12 & 610.82 \\
\bottomrule
\end{tabular}}
\caption{Statistics from the included data. \textit{Agentic Categories}: counts per category. \textit{QA Lengths}: mean word counts for questions, mean per-option length (MCQs), and open-ended answers (computed only when options are absent). \textit{Video–Audio Statistics}: duration in seconds (mean, min, max, std).}
\label{tab:subclass_stats}
\vspace{-1em}

\end{table*}

\begin{figure}
  \centering
  \includegraphics[width=\columnwidth]{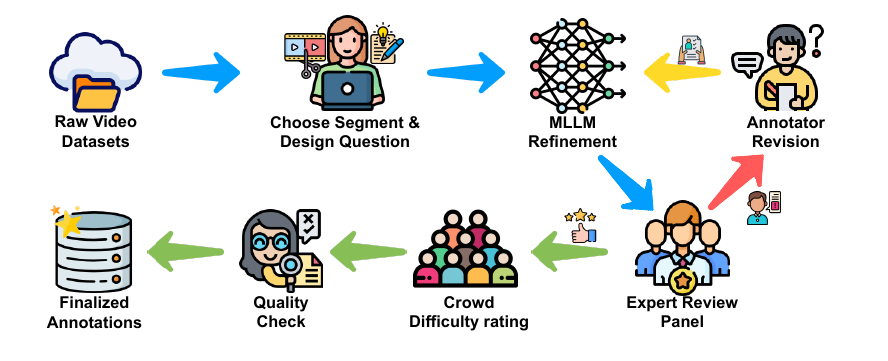}
  \caption{The framework to construct annotation sets with hybrid annotator and MLLM-as-judge quality assurance. }
\label{fig:workflow}
\vspace{-1.2em}
\end{figure}

\paragraph{Quality Assurance} 
To ensure the reliability of MAVERIX’s videos and annotations, each QA pair undergoes four checks by an expert annotator, as illustrated in Fig.~\ref{fig:workflow}: (1) linguistic validity for clear and grammatical phrasing, (2) answerability of whether the question is resolvable via the video’s audiovisual content), (3) option integrity to ensure one correct answer with plausible distractors like semantically tangent or structurally identical options, and (4) modality interdependence, using cross-modal invalidation tests from Section~\ref{sec:eval} (e.g., disabling audio or video to detect shortcuts). For open-ended questions, reviewers also confirm that rephrased variants preserve meaning without overlapping with the ground-truth wording.

\subsection{Dual‑Format Evaluation}
\label{sec:dual_eval}

\noindent\textbf{Eight‑Option MCQs.}  
Each of the 2556 questions offers one ground‑truth answer and seven carefully crafted distractors.  Annotators design distractors that remain semantically consistent with the clip yet differ in key audiovisual details, forcing models to discriminate subtle cross‑modal cues.  Expanding to eight options lowers random accuracy to \SI{12.5}{\percent}, yielding finer score resolution.  Initial difficulty labels from Gemini and GPT‑4o are later re‑calibrated by crowd workers to align machine estimates with human perception.

\noindent\textbf{Open‑Ended Generation.}  
Every clip is paired with at least one semantically unique free-form prompt, ranging from causal explanation to future prediction, requiring natural‑language output. For each unique question, we paraphrase the question and prompt the model again to test its robustness against paraphrasing. The open-ended responses are graded by GPT‑4o on a five‑factor rubric covering factual correctness, attention to detail, contextual grounding, temporal coherence, and paraphrase robustness. 

Together, the MCQ and generation tracks marry \emph{scalability}, through high-throughput accuracy metrics, with \emph{depth}, by exposing weaknesses in explanation quality.

\subsection{MAVERIX Statistics}
\label{sec:design}

MAVERIX comprises a diverse set of videos spanning 155 real-world scenarios across evaluation dimensions including agentic categories, topic domains, sub-topic domains, audio categories, duration, and difficulty. These are complemented by hierarchies over topics (e.g., travel, technology), video categories (e.g., documentaries, vlogs), and multimodal abilities (e.g., temporal reasoning, emotion recognition), supporting broad and balanced evaluation. Each question is also tagged with a difficulty level judged by human annotators. A detailed breakdown of these splits is provided in Tab. ~\ref{tab:subclass_stats}.

The dataset contains 105.8 hours of video footage, with durations ranging from 10 seconds to 63 minutes. Videos are distributed across three duration categories: 16.8\% short clips (\textless1 minute) for rapid context-switching understanding, 75.7\% medium-length videos (1-10 minutes) for sustained understanding, and 7.4\% long-form content (10-65 minutes) for testing temporal coherence. The average video length is 352.63 seconds, with an std. of 610.82, suggesting that the included medias have a diverse distribution in lengths. Constructing a audiovisual benchmark with diverse media lengths reflects real-world use cases for MLLMs and sufficiently challenges their ability to generalize. 

Each video is paired with 3 to 4 questions on average with 2,556 in total, including 852 eight-option multiple-choice questions (MCQs) and 1704 open-ended prompts, with examples shown in Fig.~\ref{fig:example}. Questions span over the evaluation dimensions to ensure a thorough evaluation. To mitigate positional bias, answer labels are uniformly redistributed across options. 

\section{Experiments}
\subsection{Evaluation Protocol}
\label{sec:eval}

MAVERIX adopts a dual evaluation framework to assess multimodal LLMs (MLLMs) through eight-option multiple-choice questions (MCQs) and open-ended response generation.

Evaluation is conducted under two settings: localized, where models access only the \textit{timestamped video segment} relevant to each question, and global (MAVERIX-Long), where the full-length video is provided. The localized setting limits the context to the specific temporal window required for understanding, whereas providing the full videos demands the models to localize the required information from the haystack of frames. 

For MCQs, we report both split-specific and overall accuracy, with answer choices uniformly distributed across positions (A-H) to reduce positional bias. Open-ended responses are evaluated using a GPT-4o scoring pipeline, adapted from Video ChatGPT~\cite{Maaz2023VideoChatGPT}, which assesses the output in five dimensions on a scale of 0-5. The results are aggregated across modalities (Tab.~\ref{tab:modality_accuracy}), with separate analyses for easy, medium, and hard videos to diagnose comprehension limitations in Tab.~\ref{tab:mcq_difficulty}. We also report the model's open-ended response qualities and token counts in Tab.~\ref{tab:open_ended_scores_updated}. Our proposed evaluation protocol ensures reproducibility while addressing modality interdependence and human performance baselines.

\subsection{Baselines}
We evaluate MAVERIX on a diverse suite of 17 MLLMs, encompassing both proprietary and open-source models, to assess their ability to reason over intertwined audiovisual modalities. Proprietary models include Gemini 2.0-FL~\cite{comanici2025gemini25pushingfrontier}, Gemini 2.5-FL~\cite{comanici2025gemini25pushingfrontier}, GPT-4o~\cite{openai2024gpt4ocard}, Grok4~\cite{grok4}, Claude Sonnet 3.5~\cite{claude3.5}, Nova-Lite~\cite{nova}, and Nova-Pro~\cite{nova}. While open source representatives feature Ola~\cite{liu2025olapushingfrontiersomnimodal}, EgoGPT~\cite{yang2025egolifeegocentriclifeassistant}, VITA 1.5~\cite{fu2025vita15gpt4olevelrealtime}, Qwen 2.5 Omni~\cite{xu2025qwen25omnitechnicalreport}, InternVL2~\cite{chen2024internvl}, Qwen2.5-VL~\cite{Qwen2.5-VL}, LLaVA-OneVision~\cite{li2024llava}, DeepSeek-VL2-Small~\cite{wu2024deepseekvl2}. Among them, Ola, EgoGPT, VITA 1.5, Qwen 2.5 Omni and Gemini are equipped with native audiovisual processing, enabling direct ingestion of raw video-audio streams. For the tested models, we maximize temporal resolution by sampling frames at their maximum supported rates. However, most architectures, including GPT-4o and LLaVA-OneVision, require transcribed subtitles as text proxies for audio. To standardize inputs, we preprocess all videos using Whisper-v3\cite{radford2022whisper} to extract time-synced subtitles, with the timestamps provided to the evaluated models.

\begin{table*}[t]
\centering
\small
\caption{Multimodal gains across models on MCQs (measured in \% accuracy). A, V, and S denote the Audio, Video, and Subtitle modalities, respectively. \textit{Best-Uni} $=\max(\text{A},\text{S},\text{V})$; \textit{Best-Multi} $=\max(\text{V+S},\text{V+A})$; $\Delta$\textit{Multi} $=$ \textit{Best-Multi} $-$ \textit{Best-Uni}.}
\setlength{\tabcolsep}{4pt}
\renewcommand{\arraystretch}{1.07}
\resizebox{0.9\textwidth}{!}{%
\begin{tabular}{
l
c
>{\footnotesize}c
>{\footnotesize}c
>{\footnotesize}c
S[table-format=2.1] 
S[table-format=2.1] 
S[table-format=2.1] 
S[table-format=2.1] 
|
S[table-format=2.1] 
S[table-format=2.1] 
S[table-format=2.1] 
|
r                 
}
\toprule
\multirow{2}{*}{\textbf{Model}} &
\multirow{2}{*}{\textbf{\begin{tabular}{@{}c@{}}Audio\\Support\end{tabular}}} &
\multirow{2}{*}{\textbf{Size}} &
\multirow{2}{*}{\textbf{Arch}} &
\multirow{2}{*}{\textbf{Recipe}} &
\multicolumn{4}{c}{\textbf{Unimodal Analysis}} &
\multicolumn{4}{c}{\textbf{Multimodal Performance}} \\
\cmidrule(r{0pt}){6-9}\cmidrule(l{0.25em}){10-13}
& & & & & \textbf{A} & \textbf{S} & \textbf{V} & \textbf{Best-Uni} & \textbf{V+S} & \textbf{V+A} & \textbf{Best-Multi} & \textbf{$\Delta$Multi} \\
\midrule
\textbf{Human}                  &      & \multicolumn{1}{c}{--} & \multicolumn{1}{c}{--} & \multicolumn{1}{c}{--}
& 44.3 & 41.7 & {\bfseries 81.4} & {\bfseries 81.4}
& {\bfseries 86.4} & {\bfseries 92.8} & {\bfseries 92.8} & \deltacol{+11.4} \\
\midrule
\textbf{EgoGPT-7B}              & \checkmark & 7B  & Dual-Tower & SFT
& 29.9 & 43.0 & 45.2 & 45.2
& 55.0 & 45.2 & 55.0 & \deltacol{+9.8} \\
\textbf{Ola-7B}                 & \checkmark & 7B  & Tri-Tower & SFT
& {\bfseries 49.4} & 43.9 & 37.6 & 49.5
& 49.6 & 53.1 & 53.1 & \deltacol{+3.6} \\
\textbf{VITA 1.5}               & \checkmark & 7B  & Dual-Tower  & SFT
& 32.4 & 43.4 & 20.2 & 43.5
& 22.3 & 18.5 & 22.3 & \deltacol{-21.2} \\
\textbf{Qwen 2.5 Omni}          & \checkmark & 7B  & Dual-Tower  & SFT+RL
& 46.5 & 41.4 & 35.4 & 46.5
& 57.9 & 49.5 & 57.9 & \deltacol{+11.4} \\
\textbf{Qwen-2-VL}              &      & 7B  & ViT-LLM  & SFT
& \multicolumn{1}{c}{--} & 43.0 & 48.0 & 48.0
& 57.5 & \multicolumn{1}{c}{--} & 57.5 & \deltacol{+9.5} \\
\textbf{Qwen-2.5-VL}            &      & 7B  & ViT-LLM  & SFT
& \multicolumn{1}{c}{--} & 40.3 & 46.9 & 46.9
& 55.3 & \multicolumn{1}{c}{--} & 55.3 & \deltacol{+8.4} \\
\textbf{InternVL2}              &      & 8B  & ViT-LLM  & SFT
& \multicolumn{1}{c}{--} & 24.1 & 26.3 & 26.3
& 33.1 & \multicolumn{1}{c}{--} & 33.1 & \deltacol{+6.8} \\
\textbf{LLaVA-OneVision}        &      & 7B  & SigLIP-LLM  & SFT
& \multicolumn{1}{c}{--} & 44.5 & 46.8 & 46.8
& 55.6 & \multicolumn{1}{c}{--} & 55.6 & \deltacol{+8.8} \\
\textbf{DeepSeekVL2-small}      &      & 2.8B  & Hybrid Enc.-MoE & SFT
& \multicolumn{1}{c}{--} & 34.3 & 33.2 & 34.3
& 42.4 & \multicolumn{1}{c}{--} & 42.4 & \deltacol{+8.1} \\
\midrule
\textbf{Gemini 2.0-FL}  & \checkmark & \multicolumn{1}{c}{--} & \multicolumn{1}{c}{--} & \multicolumn{1}{c}{--}
& 43.8 & 38.0 & 42.1 & 43.8
& 41.1 & 50.2 & 50.2 & \deltacol{+6.4} \\
\textbf{Gemini 2.5-FL}  & \checkmark & \multicolumn{1}{c}{--} & \multicolumn{1}{c}{--} & \multicolumn{1}{c}{--}
& 44.8 & 47.7 & 48.8 & 48.8
& 56.7 & 54.7 & 56.7 & \deltacol{+7.9} \\
\textbf{Claude Sonnet 3.5}      &      & \multicolumn{1}{c}{--} & \multicolumn{1}{c}{--} & \multicolumn{1}{c}{--}
& \multicolumn{1}{c}{--} & 55.0 & 42.0 & 55.0
& {\bfseries 64.1} & \multicolumn{1}{c}{--} & {\bfseries 64.1} & \deltacol{+9.1} \\
\textbf{GPT-4o}                 &      & \multicolumn{1}{c}{--} & \multicolumn{1}{c}{--} & \multicolumn{1}{c}{--}
& \multicolumn{1}{c}{--} & {\bfseries 55.3} & 54.3 & {\bfseries 55.3}
& 64.0 & \multicolumn{1}{c}{--} & 64.0 & \deltacol{+8.7} \\
\textbf{Grok 4}                 &      & \multicolumn{1}{c}{--} & \multicolumn{1}{c}{--} & \multicolumn{1}{c}{--}
& \multicolumn{1}{c}{--} & 41.8 & 54.5 & 54.5
& 59.4 & \multicolumn{1}{c}{--} & 59.4 & \deltacol{+4.9} \\
\textbf{GPT-4o-mini}            &      & \multicolumn{1}{c}{--} & \multicolumn{1}{c}{--} & \multicolumn{1}{c}{--}
& \multicolumn{1}{c}{--} & 45.4 & 35.5 & 45.4
& 50.0 & \multicolumn{1}{c}{--} & 50.0 & \deltacol{+4.6} \\
\textbf{NOVA-Lite}              &      & \multicolumn{1}{c}{--} & \multicolumn{1}{c}{--} & \multicolumn{1}{c}{--}
& \multicolumn{1}{c}{--} & 40.4 & 39.7 & 40.4
& 51.0 & \multicolumn{1}{c}{--} & 51.0 & \deltacol{+10.6} \\
\textbf{NOVA-Pro}               &      & \multicolumn{1}{c}{--} & \multicolumn{1}{c}{--} & \multicolumn{1}{c}{--}
& \multicolumn{1}{c}{--} & 46.6 & 45.4 & 46.6
& 55.8 & \multicolumn{1}{c}{--} & 55.8 & \deltacol{+9.2} \\
\bottomrule
\end{tabular}}
\label{tab:modality_accuracy}
\end{table*}

All models receive inputs in the unified format [video frames, subtitles, question], with frames uniformly sampled at their maximum supported context window. For audio-incapable models, subtitles replace raw audio tracks, while the audio-supported models additionally process synchronized audio-video pairs. We employ a standardized prompt template across models, ensuring fairness by eliminating instructional biases. This setup isolates modality interdependence as the critical challenge: models must synthesize potentially asynchronous audiovisual cues, such as the startled expression of a character with an auditory context like an off-screen crash to match human-like understanding.

\definecolor{highlight}{RGB}{255, 255, 204} 

\definecolor{videoonly}{RGB}{230, 242, 255} 
\definecolor{audioonly}{RGB}{255, 242, 204} 
\definecolor{textonly}{RGB}{232, 245, 233} 
\definecolor{videosub}{RGB}{255, 224, 230} 
\definecolor{videoaudio}{RGB}{217, 234, 211} 
\definecolor{videoaudiosub}{RGB}{242, 220, 219} 
\definecolor{categoryblue}{RGB}{255, 255, 255} 
\definecolor{lightblue}{RGB}{248, 250, 252} 
\definecolor{lightgreen}{RGB}{239, 243, 234} 
\definecolor{lightyellow}{RGB}{255, 253, 240} 
\definecolor{lightpink}{RGB}{255, 227, 227} 
\definecolor{lightgray}{RGB}{255, 255, 255} 

\newcommand{\up}[1]{\textsuperscript{\textcolor{teal}{\scriptsize\,#1$\uparrow$}}}
\newcommand{\down}[1]{\textsuperscript{\textcolor{red}{\scriptsize\,#1$\downarrow$}}}

\begin{table}[t]
\centering
\caption{Difficulty-wise MCQ accuracy (\%) for audio-enabled models. AV cells show $\Delta$ vs A.}
\label{tab:mcq_difficulty}
\setlength{\tabcolsep}{2pt}
\renewcommand{\arraystretch}{1.01}
\resizebox{0.48\textwidth}{!}{%
\begin{tabular}{l | ccc | ccc | ccc}
\toprule
\textbf{Model} & \multicolumn{3}{c|}{\textbf{Easy}} & \multicolumn{3}{c|}{\textbf{Medium}} & \multicolumn{3}{c}{\textbf{Hard}} \\

               & A & V & AV & A & V & AV & A & V & AV \\
\midrule
Human & 46.4 & 84.7 & 93.4\up{47.0} & 44.8 & 81.4 & 92.5\up{47.7} & 38.5 & 73.9 & 92.1\up{53.6} \\
\midrule
EgoGPT & 29.4 & 50.2 & 50.2\up{20.8} & 32.3 & 44.3 & 44.3\up{12.0} & 25.2 & 36.4 & 36.4\up{11.2} \\
Ola-7B & 54.1 & 36.9 & 57.1\up{3.0}  & 48.4 & 41.0 & 54.1\up{5.7}  & 41.7 & 30.5 & 41.1\down{0.6} \\
VITA 1.5 & 33.0 & 21.3 & 20.1\down{12.9} & 34.8 & 19.8 & 18.5\down{16.3} & 25.2 & 18.5 & 14.6\down{10.6} \\
Qwen-2-Omni & 50.6 & 39.9 & 52.1\up{1.5} & 44.6 & 34.8 & 48.8\up{4.2} & 41.5 & 25.9 & 45.2\up{3.7} \\
\midrule
Gemini 2.0-FL & 43.8 & 45.3 & 57.7\up{13.9} & 44.0 & 42.7 & 48.9\up{4.9} & 32.5 & 33.1 & 36.4\up{3.9} \\
Gemini 2.5-FL & 47.7 & 48.3 & 59.8\up{12.1} & 44.8 & 53.5 & 54.9\up{10.1} & 37.7 & 37.7 & 43.0\up{5.3} \\
\bottomrule
\end{tabular}}
\vspace{-1em}
\end{table}

\section{Model analysis with \textsc{Maverix}}

This section discusses model behaviour along five dimensions: multimodal gains across modal deisgns, training‑recipe variation, agentic ability relative to humans, temporal‑horizon sensitivity, and perceived question difficulty, using the benchmark set. 

\noindent\textbf{Multimodal Gains Across Model Capacity and Architecture.}
Tab.~\ref{tab:modality_accuracy} summarizes performance across five architectural families and multiple model sizes. Best unimodal accuracies span $\sim$26–55\%, and most models show gains when additional modalities are provided. Many systems follow an audiovisual encoder$+$MLP$+$LLM design; among these, Ola, EgoGPT, and Qwen~2.5~Omni, which incorporate Whisper~v3 for audio, generally do not regress and often improve relative to their strongest unimodal scores, whereas VITA~1.5 shows a regression with audio–visual input. The lightweight Gemini~2.0-FL and 2.5-FL variants also improve with multimodal inputs. For video-text models, adding a modality yields consistent gains; Qwen2-VL, Nova-Lite, Nova-Pro, and Claude~3.5~Sonnet improve by roughly $\sim$10\%. Despite these gains, a sizable gap to human performance remains, suggesting that current models under-utilize cross-modal cues. This is evident when contrasting V$+$S and V$+$A: several models (e.g., Gemini~2.5-FL, EgoGPT, VITA~1.5, and Qwen~2.5~Omni) score lower with V$+$A than with V$+$S, indicating missed auditory details or limitations in audio-video fusion.

\begin{table}[t]
\centering
\tiny
\caption{Accuracy (\%) on AV inputs across taxonomy and audio categories (single-column).}
\label{tab:combined_taxonomy_audio_av}
\setlength{\tabcolsep}{5pt}
\renewcommand{\arraystretch}{1.12}

\begin{minipage}{0.45\textwidth}
\centering
\resizebox{\linewidth}{!}{%
\begin{tabular}{lcccc}
\toprule
\multicolumn{5}{c}{\textbf{Taxonomy (AV)}}\\
\cmidrule(lr){2-5}
\textbf{Models} & \textbf{Social} & \textbf{Ego-centric} & \textbf{Sports} & \textbf{Gaming} \\
\midrule
Human         & 92.7 & 95.2 & 82.3 & 74.5 \\
EgoGPT        & 34.3 & 45.8 & 42.9 & 40.8 \\
Ola           & 52.5 & 57.6 & 52.9 & 46.6 \\
VITA 1.5      & 21.2 &  6.8 & 16.8 & 12.6 \\
Qwen 2.5 Omni & 47.5 & 42.4 & 49.6 & 41.7 \\
Gemini 2.0-FL & 46.5 & 32.3 & 48.7 & 47.6 \\
Gemini 2.5-FL & 47.1 & 39.0 & 54.3 & 61.9 \\
\end{tabular}}
\end{minipage}


\begin{minipage}{0.45\textwidth}
\centering
\resizebox{\linewidth}{!}{%
\begin{tabular}{lccccc}
\toprule
\multicolumn{6}{c}{\textbf{Audio Category (AV)}}\\
\cmidrule(lr){2-6}
\textbf{Models} & \textbf{Natural} & \textbf{Speech} & \textbf{Music} & \textbf{Artificial} & \textbf{Mixed} \\
\midrule
Human         & 88.4 & 94.1 & 90.3 & 89.6 & 92.8 \\
EgoGPT        & 41.2 & 44.2 & 46.2 & 48.0 & 47.3 \\
Ola           & 48.5 & 52.0 & 38.5 & 58.0 & 56.5 \\
VITA 1.5      & 29.4 & 17.0 &  7.7 & 26.0 & 17.6 \\
Qwen 2.5 Omni & 52.9 & 47.8 & 53.8 & 48.0 & 51.5 \\
Gemini 2.0-FL & 47.1 & 51.1 & 38.5 & 54.0 & 49.6 \\
Gemini 2.5-FL & 51.5 & 56.7 & 50.0 & 52.0 & 53.1 \\
\bottomrule
\end{tabular}}
\end{minipage}
\vspace{-1em}
\end{table}

\noindent\textbf{Training-Recipe Variants: SFT, RL, and Data Composition.}
We assess open-source recipes built on identical backbones to isolate curriculum effects. Omni-modal models such as Ola and Qwen 2.5 Omni use an image-text warmup followed by separate alignment for the audio modality, and they typically improve on audiovisual evaluations; reinforcement learning appears to further increase the multimodal gains. VITA 1.5, on the other hand, emphasizes alignment to the video modality during training, which may bias attention toward the visual stream and results in regression when subtitles and audio are added.

Video-text models follow a more streamlined path: initial pretraining on image-caption pairs to align images with text, then instruction tuning or long chain-of-thought data for fine-tuning. Aside from minor architectural differences, they vary primarily in data curation and sources. Qwen 2.5 VL uniquely includes chain-of-thought data during fine-tuning to encourage explicit reasoning and stronger multimodal synthesis. While its overall performance is strong, the relative gain from multimodal inputs appears similar to its counterpart without chain-of-thought fine-tuning, which may point to reward hacking during SFT and warrants further study.

\begin{table}[t]
\centering
\caption{Open-ended response correctness scores (out of 5) as judged by GPT, reported per modality. For models without native audio support, scores for Subtitle (S) and Subtitle+Video (SV) are shown instead of Audio (A) and Audio+Video (AV). Response token length statistics are analyzed separately.}
\label{tab:open_ended_scores_updated}
\setlength{\tabcolsep}{1pt}
\renewcommand{\arraystretch}{0.95}
\resizebox{0.45\textwidth}{!}{%
\begin{tabular}{l c |ccc|ccc}
\toprule
\multirow{2}{*}{\textbf{Model}} & \multirow{2}{*}{\textbf{\begin{tabular}{@{}c@{}}Audio\\Support\end{tabular}}} & \multicolumn{3}{c|}{\textbf{Judged Score}} & \multicolumn{3}{c}{\textbf{Token Length}} \\
\cmidrule(lr){3-5} \cmidrule(lr){6-8}
& & \textbf{A/S} & \textbf{V} & \textbf{AV/SV} & \textbf{Avg} & \textbf{Max} & \textbf{Std} \\
\midrule
Human                      &            & 1.7 & 2.6 & 3.4\up{0.8} & 16.1 & 140   & 12.7 \\
\midrule
EgoGPT                     & \checkmark & 0.7 & 1.2 & 1.3\up{0.1} & 9.4  & 112   & 17.8 \\
Ola-7B                     & \checkmark & 1.4 & 0.9 & 1.5\up{0.1} & 17.9 & 71    & 15.4 \\
VITA 1.5                   & \checkmark & 0.8 & 0.7 & 0.5\down{0.3} & 55.3 & 273   & 29.1 \\
Qwen-2-Omni                & \checkmark & 1.0 & 1.2 & 1.2\down{0.0} & 47.0 & 178   & 26.5 \\
Qwen-2-VL                  &            & 1.2 & 1.4 & 1.6\up{0.2} & 22.9 & 72    & 15.7 \\
Qwen-2.5-VL                &            & 1.1 & 1.4 & 1.7\up{0.3} & 52.9 & 86    & 17.7 \\
InternVL2                  &            & 0.9 & 0.9 & 1.1\up{0.2} & 24.0 & 95    & 17.6 \\
DeepSeek-VL2               &            & 1.2 & 1.0 & 1.4\up{0.2} & 27.4 & 512   & 61.0 \\
LLaVA-OneVision            &            & 1.3 & 1.4 & 1.6\up{0.2} & 18.1 & 70    & 9.3  \\
\midrule
Gemini 2.0-FL      & \checkmark & 1.4 & 1.6 & 1.9\up{0.3} & 20.9 & 103   & 14.6 \\
Gemini 2.5-FL      & \checkmark & 1.4 & 1.4 & 1.9\up{0.5} & 32.6 & 112   & 20.7 \\
Claude Sonnet 3.5          &            & 1.6 & 1.7 & 2.2\up{0.5} & 59.8 & 95    & 4.8  \\
GPT-4o                     &            & 1.6 & 1.4 & 2.2\up{0.6} & 50.8 & 102   & 17.9 \\
Grok-4                     &            & 1.7 & 2.0 & 2.4\up{0.4} & 131.7& 11022 & 322.3 \\
NOVA-Lite                  &            & 1.1 & 1.0 & 1.2\up{0.1} & 23.5 & 75    & 18.4 \\
NOVA-Pro                   &            & 1.1 & 1.2 & 1.5\up{0.3} & 35.4 & 78    & 20.9 \\
\bottomrule
\end{tabular}}
\vspace{-1em}
\end{table}

\noindent\textbf{Agentic Ability in Comparison with Humans.}
Tab.~\ref{tab:combined_taxonomy_audio_av} indicates that humans perform best on social and egocentric questions, with slightly lower accuracy on gaming and sports that demand domain knowledge. Model behavior is less uniform. Gemini, Qwen 2.5 Omni, and VITA 1.5 tend to be weaker on egocentric videos and comparatively stronger on domain-specific categories, while the remaining models show different per-category strengths.

Across audio categories, human accuracy is largely stable. Models trained on broader multimodal corpora such as Qwen 2.5 Omni and the Gemini family exhibit smaller fluctuations across audio types, yet many systems under-perform when music dominates. EgoGPT shows relatively strong auditory understanding, likely reflecting its use of a pretrained Whisper encoder. In contrast, Ola and VITA 1.5 that trained on smaller datasets, display larger variance across categories, with notable drops on the music category.

Taken together, these patterns suggest that coverage of first-person content and diverse audio during training may be as important as scale for robust agentic ability across domains.

\begin{figure}
  \centering
  \includegraphics[width=\columnwidth]{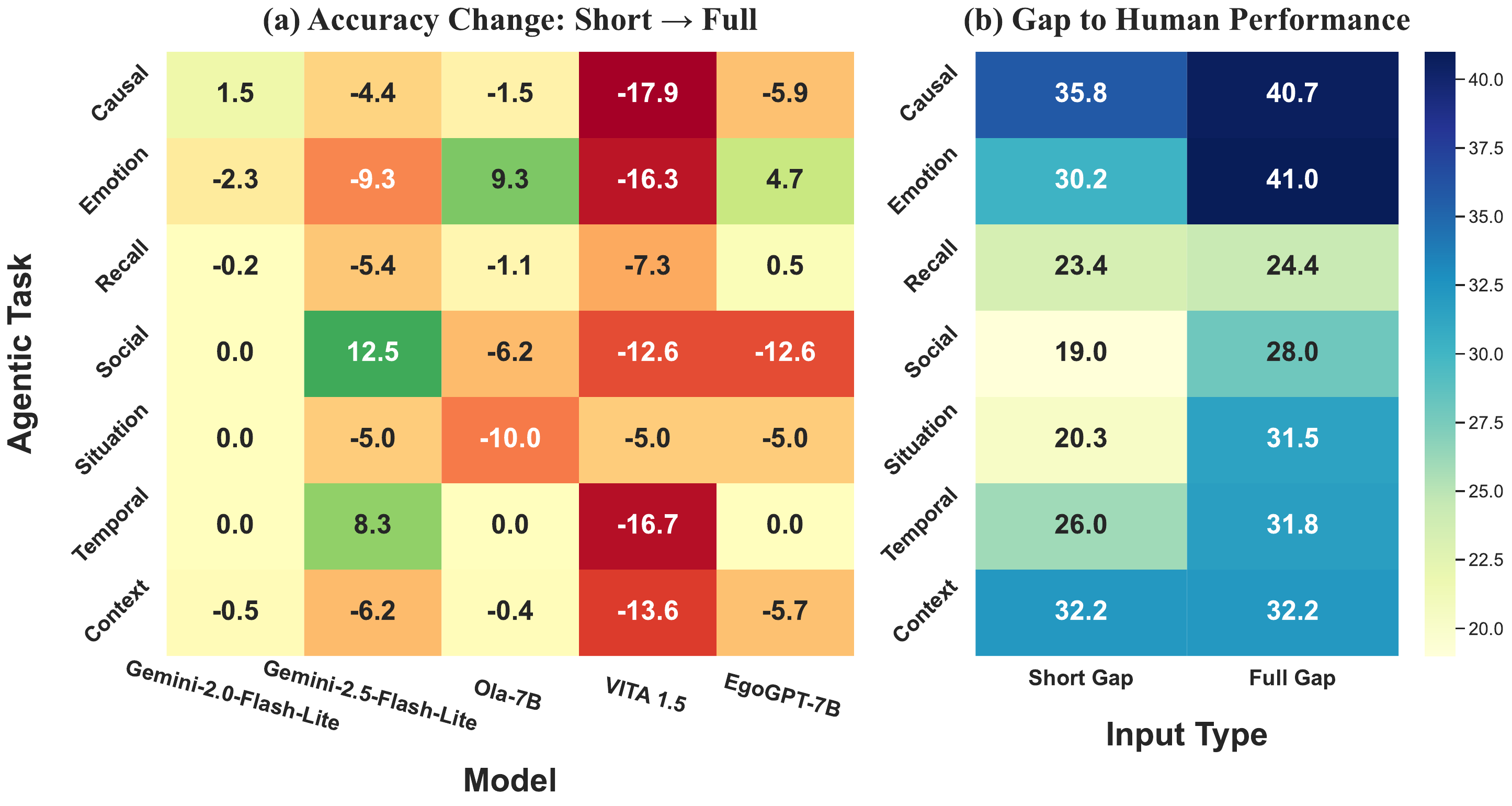}
  \caption{Impact of Video Length on Agentic Category Performances.
(a) Accuracy change (\%) from short to full-length videos across models and questions.
(b) Accuracy gap (\%) to human performance for short and full inputs.}
\label{fig:gap_heatmap}
\vspace{-1.2em}
\end{figure}

\noindent\textbf{Temporal Horizons: Long- versus Short-Clip Performance.}
As shown in Fig.~\ref{fig:gap_heatmap}, we evaluate models on short, pre-localized audiovisual clips in MAVERIX and on their full-length counterparts in MAVERIX-Long. Across models, localized clips yield higher accuracy. Among the agentic categories, the questions that depend on immediate, synchronous audiovisual cues, such as those from factual recall and near-term causal inference, show the smallest degradation. When the relevant segment is pre-localized, models can more reliably extract the necessary information.

By contrast, social relationship, emotion, and situational understanding often rely on fine-grained and sometimes asynchronous cues distributed over time. Performance drops more on long videos, reflecting challenges in localizing these signals and integrating them over extended context. Overall, a gap to human performance remains, especially for longer videos and for recognizing subtle contextual cues. These trends suggest that current MLLMs are stronger at retrieving salient, object or event level signals than at integrating evolving context and social nuance over time.

\noindent\textbf{Model and Human Perception of Difficulty.}
We analyze performance by difficulty and observe that multimodal inputs often help most on easy items, with smaller gains on hard ones, though trends vary by model. For the audio-enabled models where split statistics are available, Gemini 2.0-FL improves by $+$12.4\% on easy items and $+$3.3\% on hard items, and Gemini 2.5-FL improves by $+$11.5\% on easy items and $+$5.3\% on hard items. Qwen 2.5 Omni shows a different pattern with substantial benefit on hard items as well. These mixed results suggest that current systems leverage straightforward cross-modal cues more reliably than they integrate sparse or subtle signals in harder cases.

Although MAVERIX is designed to elicit cross-modal reasoning, some models still achieve moderate scores with a single modality, likely because many real videos contain aligned audio and visual streams that allow plausible inferences from partial evidence. GPT-4o-mini is one such example of respectable unimodal performance. Humans also benefit from aligned cues and strong priors, yet the jump from 81.4\% with video-only to 92.8\% with audiovisual highlights the value of genuine cross-modal understanding and sets a clear target for future modeling. We hope continued progress in cross-modal alignment will narrow this gap and eventually surpass the current human baseline.

\section{Related Work}

\textbf{MLLM Benchmarks.}
Early vision-language benchmarks centered on images for grounding and recognition, using captions and QA pairs~\cite{chen2015microsoft,agrawal2019nocaps,zhu2016visual7w,plummer2015flickr30k}, followed by domain-specific and knowledge-intensive settings~\cite{saikh2022scienceqa,lu2023mathvista,singh2019towards}. A-OKVQA targets external-knowledge reasoning beyond visible content~\cite{schwenk2022okvqa}. More recent efforts, including MMMU and MMMU-Pro, broaden question diversity and reading-from-image skills~\cite{yue2024mmmu,yue2024mmmupro}.

Image-only evaluation lacks temporal and acoustic context, motivating video benchmarks that probe motion, sequence, and temporal localization~\cite{li2024mvbench,patraucean2023perception,kesen2023vilma,huang2020movie,moviechat,Maaz2023VideoChatGPT,fang2025mmbench,li2024vitatecs,ning2023video,chen2024autoeval,he2024mmworld,mangalam2023egoschema,song2024moviechat}. However, most emphasize short clips and constrained domains, rely on MCQ-only protocols, and provide limited coverage of everyday social or situational reasoning. Video-MME and AV-Odyssey scale video duration but remain MCQ-only, omitting open-ended assessment~\cite{fu2024video, gong2024avodysseybenchmultimodalllms}. 
\emph{MAVERIX} elevates audio as one of the primary signals and stresses cross-modal integration as it evaluates both 8-way MCQs and open-ended responses to measure multimodal synthesis under realistic audiovisual conditions.

\noindent\textbf{Video Understanding Models.}
Contrastive pretraining on image-text data yields transferable representations and has been adapted to spatio-temporal reasoning; post-training with instruction tuning and RLHF further aligns models with human preferences~\cite{2023llavarlhf,zhai2024finetuning,lin2023learning,Qwen2-VL,Qwen2.5-VL,vlrlhf}. MoE-style routing improves scalability without linear cost growth~\cite{wu2024deepseekvl2,lin2024moe,deepseekai2025deepseekv3technicalreport,sun2024hunyuanlargeopensourcemoemodel,cai2024flextronmanyinoneflexiblelarge,liu2025surveyinferenceoptimizationtechniques}. Yet whether these advances enable human-comparable multimodal reasoning in real-world audiovisual settings remains open~\cite{testingnature,llmsreason,campbell2024understandinglimitsvisionlanguage,kazemi2024remidatasetreasoningmultiple}. Our evaluations on \emph{MAVERIX} show a substantial gap to human accuracy on MCQs, often on the order of several tens of percentage points, and highlight persistent challenges in integrating temporal, social, and auditory cues in models with different architecture and training recipes.

\section{Conclusion}
Agentic scenarios such as assisting collaborative work and navigating dynamic environments require strong audiovisual reasoning, yet these abilities remain under-assessed in recent MLLMs. We introduce \textbf{MAVERIX}, a benchmark for complex, real-world audiovisual understanding, comprising 700 videos and 2{,}556 carefully crafted, human-authored questions. The suite evaluates models with both 8-way multiple-choice and open-ended responses.

Our results indicate that multimodal inputs generally improve accuracy, but sizable gaps to human performance persist, especially for socially grounded or dynamic scenarios. Models benefit most when relevant segments are pre-localized and tend to struggle on longer videos that demand integrating subtle, asynchronous cues over time; robust audio integration also remains uneven across systems. We hope MAVERIX will guide progress toward stronger cross-modal alignment, better temporal reasoning, and more context-aware, socially intelligent models.

{
    \small
    \bibliography{aaai2026}
}


\UseRawInputEncoding

\clearpage
\appendix


\section*{MAVERIX Supplementary Material}
\addcontentsline{toc}{section}{MAVERIX Supplementary Material}

\begin{itemize}
    \item Section A: MAVERIX Release v1.0
    \item Section B: Rationale and Limitations
    \item Section C: Additional Experiments
    \begin{itemize}
        \item Section C.1: Standardized Prompt Design
        \item Section C.2: GPT-4o as Judge Criteria
        \item Section C.3: Human Performance by Categories
        \item Section C.4: Quantitative GPT Evaluations
        \item Section C.5: Performance by Categories
        \item Section C.6: Common Error Modes
    \end{itemize}
    \item Section D: Details on the Human Survey
    \item Section E: Dataset Distribution Continued

\end{itemize}

\section{A. MAVERIX Release v1.0}
\label{sec:release}

We are releasing \textbf{MAVERIX v1.0}, our proposed benchmark dataset for multimodal video-language understanding, built with support to a common benchmark platform LMMs-Eval~\cite{lmms_eval2024}. The benchmark dataset is provided as a single JSON file for ease of use and straightforward integration with existing benchmarking pipelines. For each video, the dataset includes metadata and contains multiple-choice and open-ended questions covering multiple tasks from our proposed task suite. Each task is accompanied by a set of questions designed to evaluate visual reasoning, situational awareness, and cross-modal understanding. Relevant timestamps are provided to allow precise video trimming.

We will provide all 1400 video clips used in our benchmark through HuggingFace~\cite{huggingface-hub}, with both localized and full length videos, and users can download the corresponding videos from the Ego4D website after reviewing and accepting the Ego4D license agreement. We also provide sample videos with annotations from MAVERIX. All materials are available through the project page upon acceptance. The code snippets providing main input prompts, response aggregation, judge response generations are provided as part of the supplement materials, with unit test functions for testing data corruption included. 

\vspace{1em}
\noindent\textbf{Structure:} MAVERIX v1.0 release is organized as follows:
\begin{itemize}[left=0pt]
    \item \textbf{\texttt{data/}}
    \begin{itemize}
        \item \texttt{MAVERIX\_v1\_0.json}: Contains all 2556 questions in the benchmark dataset.
        \item \texttt{vid\_only/}: Contains all muted videos for evaluations without audio modality. 
        \item \texttt{vid/}: Contains all videos for evaluations with both video and audio modality.
        \item \texttt{audio/}: Contains all audio files for evaluations with audio modality.
        \item \texttt{subtitles/}: Contains all subtitles for evaluations requiring subtitle access.
        \item \texttt{sample\_annotations/}: Given that \textbf{MAVERIX} is an evaluation benchmark, ground truth annotations are not released to the public. For review purposes, we provide ground truth annotations for select sample videos in an \texttt{.json} file.
    \end{itemize}

    \item \textbf{\texttt{src/}}
    \begin{itemize}
        \item \texttt{maverix\_benchmark.py}: A script for dataloading, processing, and evaluation functionalities.
        \item \texttt{\_default\_template\_yaml}: A file to specify the default processing template for the evaluation. The data loading path, cache directory, and evaluation modes can be specified here. 
        \item \texttt{maverix\_<mode>.yaml}: Individual setting files specifying the test settings for different modes. The currently supported modes are: 
        \mbox{\texttt{audio\_only}}, 
        \mbox{\texttt{video\_only}}, 
        \mbox{\texttt{sub\_only}}, 
        \mbox{\texttt{video\_sub}}, and 
        \mbox{\texttt{video\_audio}}.
        
        \item \textbf{Remark}: Except for the open-sourced models listed in the paper or available through LMMs-Eval\cite{zhang2024lmmsevalrealitycheckevaluation, lmms_eval2024}, all other experiments require access to proprietary models.
    \end{itemize}
\end{itemize}

\vspace{1em}
\noindent\textbf{Documentation:}
\begin{itemize}[left=0pt]
    \item We provide a comprehensive datasheet explaining the benchmark dataset’s purpose and intended usage.
\end{itemize}

\vspace{1em}
\noindent\textbf{License:}
\begin{itemize}[left=0pt]
    \item MAVERIX will be made publicly available under the MIT License. Do note that Ego4D videos are publicly available under the Ego4D License.
\end{itemize}

\vspace{1em}
\noindent\textbf{Versioning and Updates:}
\begin{itemize}[left=0pt]
    \item We will maintain MAVERIX, with all updates and new versions announced publicly.
\end{itemize}

\section{B. Rationale and limitations}
\label{sec:rationale}

\paragraph{Broader Impact.}
\label{sec:broader}
Robust audiovisual understanding is critical for the next generation of agentic systems that interact with humans and environments in real-time. Benchmarks like MAVERIX can help guide the development of models capable of supporting complex tasks such as collaborative decision-making, social interaction understanding, embodied navigation, and context-aware assistance. By evaluating models on realistic, multimodal scenarios, MAVERIX encourages progress toward AI systems that can reason beyond static or unimodal inputs, making them better suited for applications like personal assistants, robotics, and human-computer interaction.

However, improved performance on MAVERIX may not directly translate to safe or ethical deployment. Many categories in MAVERIX, such as social sentiment and situational awareness, involve subjective interpretation and context-sensitive judgment. Without careful handling, these capabilities risk reinforcing biases, misunderstanding cultural cues, or misinterpreting social dynamics, especially when applied across diverse real-world contexts. We encourage users of MAVERIX to consider these risks and to accompany performance improvements with a thorough evaluation of fairness, robustness, and societal impact.

\paragraph{Limitations.}
\label{sec:limitation}

While MAVERIX advances the evaluation of multimodal reasoning, it has several limitations. First, the benchmark focuses on short- to medium-length video segments, which may not fully capture the long-horizon dependencies found in extended real-world interactions. Second, although MAVERIX emphasizes modality interdependence, complete isolation of modalities is challenging, and some tasks may still be partially solvable through strong unimodal priors. Third, certain categories, such as social interaction, show lower human agreement due to the inherent subjectivity of emotional and interpersonal interpretation, which may limit the upper bound of achievable accuracy. Finally, while MAVERIX supports open-ended evaluation using GPT-4o as an automatic judge, automated scoring of generative responses remains imperfect and may introduce noise or bias in assessments.

Despite these limitations, MAVERIX offers a step toward a more comprehensive and realistic evaluation of multimodal LLMs, with the goal of driving future work in building models capable of deeper, contextually grounded, and socially aware reasoning.

\section{C. Additional experiments}
\label{sec:exp}
This section presents extended experimental results to further analyze model performance across various dimensions of the benchmark. In addition to model evaluation, we provide human performance baselines segmented by difficulty levels, input modalities, and agentic task categories for direct comparison.

\subsection{C.1 Standardized Prompt Design}
\label{subsec:prompt}
To ensure fairness in evaluation, we adopt a standardized prompt format across all models. This eliminates potential biases introduced by prompt engineering and ensures that differences in performance are attributed to model capabilities rather than variations in input phrasing. For tasks requiring multiple modalities, the prompts incorporate the modalities in the inputs. In multiple-choice settings, the prompt is presented as follows:
\begin{mdframed}
\small\texttt{
\textcolor{RoyalBlue}{[images]} 
+
\textcolor{SpringGreen}{[audio]} (if available)
This video's subtitles are listed below: OR No subtitles available.
\textcolor{BurntOrange}{[Subtitles]}
Select the best answer to the following multiple-choice question based on the video and the subtitles. Respond with only the letter (A, B, C, D, E, F, G, or H) of the correct option. 
\textcolor{Apricot}{[Question]} 
The best answer is:
}
\end{mdframed}

For open-ended questions, the following prompt is used:

\begin{mdframed}
\small\texttt{
\textcolor{RoyalBlue}{[images]} 
+
\textcolor{SpringGreen}{[audio]} (if available)
This video's subtitles are listed below: OR No subtitles available. 
\textcolor{BurntOrange}{[Subtitles]}
Select the best answer to the following open-ended question based on the video and the subtitles. 
\textcolor{Apricot}{[Question]} 
The best answer is:
}
\end{mdframed}

\subsection{C.2 GPT-4o as Judge Criteria}
\label{subsec:prompt_gpt}
For evaluating the open-ended answer qualities, we divide the evaluations in 4 different aspects: correctness, level of details, context understanding, and temporal consistency. We prompt the judge model as follows for correctness:

\begin{mdframed}[linewidth=0.5pt,backgroundcolor=gray!3]
\begin{lstlisting}[style=jsonprompt]
[
  {
    "role": "system",
    "content": "You are an intelligent chatbot designed for evaluating the factual accuracy of generative outputs for video-based question-answer pairs. Your task is to compare the predicted answer with the correct answer and judge factual consistency.\n\nINSTRUCTIONS:\n- Focus on factual consistency; avoid misinterpretations or misinformation.\n- Ensure the prediction aligns with the video's content.\n- Treat synonyms or paraphrases as valid.\n- Output a single score reflecting factual accuracy."
  },
  {
    "role": "user",
    "content": "Please evaluate the following video-based question-answer pair:\n\nQuestion: QUESTION \nCorrect Answer: REFERENCE_ANSWER \nPredicted Answer: MODEL_ANSWER\nReturn only a Python dict string like {'score': 4}, where 'score' is an INTEGER from 0 (lowest) to 5 (highest). Do not include any other text."
  }
]
\end{lstlisting}
\end{mdframed}

The template prompt for level of detail orientation:

\begin{mdframed}[linewidth=0.5pt,backgroundcolor=gray!3]
\begin{lstlisting}[style=jsonprompt]
[
  {
    "role": "system",
    "content": "You are an intelligent chatbot designed for evaluating the detail orientation of generative outputs for video-based question-answer pairs. Your task is to compare the predicted answer with the correct answer and judge its level of detail, considering both completeness and specificity.\n\nINSTRUCTIONS:\n- Check whether the prediction covers all major points; do not omit key aspects.\n- Prefer specific, grounded details over generic statements; tie information to concrete elements of the video.\n- Treat synonyms or paraphrases as valid.\n- Output a single score reflecting overall detail orientation (completeness + specificity)."
  },
  {
    "role": "user",
    "content": "Please evaluate the following video-based question-answer pair:\n\nQuestion: QUESTION\nCorrect Answer: REFERENCE_ANSWER\nPredicted Answer: MODEL_ANSWER\n\nReturn only a Python dict string like {'score': 4}, where 'score' is an INTEGER from 0 (lowest detail) to 5 (highest detail). Do not include any other text."
  }
]
\end{lstlisting}
\end{mdframed}

The template prompt for level of context understanding:

\begin{mdframed}[linewidth=0.5pt,backgroundcolor=gray!3]
\begin{lstlisting}[style=jsonprompt]
[
  {
    "role": "system",
    "content": "You are an intelligent chatbot designed for evaluating the contextual understanding of generative outputs for video-based question-answer pairs. Your task is to compare the predicted answer with the correct answer and judge whether the response aligns with the overall context of the video.\n\nINSTRUCTIONS:\n- Check that the prediction aligns with the video's context; it should not introduce out-of-context or contradictory information.\n- Ensure the prediction captures the main themes and sentiments of the video.\n- Treat synonyms or paraphrases as valid.\n- Output a single score reflecting overall contextual understanding."
  },
  {
    "role": "user",
    "content": "Please evaluate the following video-based question-answer pair:\n\nQuestion: QUESTION\nCorrect Answer: REFERENCE_ANSWER\nPredicted Answer: MODEL_ANSWER\n\nReturn only a Python dict string like {'score': 4}, where 'score' is an INTEGER from 0 (lowest alignment) to 5 (highest alignment). Do not include any other text."
  }
]
\end{lstlisting}
\end{mdframed}

The template prompt for level of temporal consistency:
\begin{mdframed}[linewidth=0.5pt,backgroundcolor=gray!3]
\begin{lstlisting}[style=jsonprompt]
[
  {
    "role": "system",
    "content": "You are an intelligent chatbot designed for evaluating the temporal understanding of generative outputs for video-based question-answer pairs. Your task is to compare the predicted answer with the correct answer and judge whether the response preserves the temporal sequence of events.\n\nINSTRUCTIONS:\n- Verify temporal consistency; the prediction should reflect the order and timing of events as presented in the video.\n- Accept synonyms or paraphrases only if the temporal order is maintained.\n- Output a single score reflecting overall temporal accuracy."
  },
  {
    "role": "user",
    "content": "Please evaluate the following video-based question-answer pair:\n\nQuestion: QUESTION\nCorrect Answer: REFERENCE_ANSWER\nPredicted Answer: MODEL_ANSWER\\n\nReturn only a Python dict string like {'score': 4}, where 'score' is an INTEGER from 0 (lowest temporal consistency) to 5 (highest). Do not include any other text."
  }
]
\end{lstlisting}
\end{mdframed}




\subsection{C.3 Human Performance by Categories}
\label{subsec:category}

We report performance across distinct agentic task categories in Tab.~\ref{tab:human_performance} to showcase the strengths and weaknesses of the human study participants within the defined contexts. We note that these statistics may be biased due to the composition of the participant pool, and further details on the recruitment process can be found in the Questionnaire Survey section.

\begin{table}[ht]
\centering
\setlength{\tabcolsep}{1pt}
\resizebox{0.45\textwidth}{!}{
\begin{tabular}{llccc}
\toprule
\textbf{Modality} & \textbf{Task} & \textbf{Worst (\%)} & \textbf{Best (\%)} & \textbf{Average (\%)} \\
\midrule
\multirow{7}{*}{Visual}
& Social               & 66.71 & 94.72 & 79.31 \\
& Sports               & 28.91 & 93.84 & 74.34 \\
& Information Querying & 77.84 & 95.67 & 89.83 \\
& Sentiment            & 28.91 & 92.43 & 68.96 \\
& Shopping             & 51.62 & 94.11 & 83.37 \\
& Gaming               & 54.20 & 96.34 & 84.18 \\
\cmidrule{2-5}
& \textbf{Overall}     & \textbf{28.91} & \textbf{96.34} & \textbf{80} \\
\midrule
\multirow{7}{*}{Audio}
& Social               & 22.10 & 53.04 & 32.05 \\
& Sports               & 26.52 & 56.83 & 44.51 \\
& Information Querying & 30.60 & 48.63 & 44.12 \\
& Sentiment            & 28.42 & 53.04 & 46.89 \\
& Shopping             & 30.60 & 79.56 & 58.48 \\
& Gaming               & 33.15 & 33.15 & 33.15 \\
\cmidrule{2-5}
& \textbf{Overall}     & \textbf{22.10} & \textbf{79.56} & \textbf{43.2} \\
\midrule
\multirow{7}{*}{Both (A+V)}
& Social               & 78    & 95.81 & 92.86 \\
& Sports               & 50.15 & 97.20 & 94.71 \\
& Information Querying & 85.80 & 94.91 & 94.77 \\
& Sentiment            & 96.28 & 96.70 & 96.87 \\
& Shopping             & 70.20 & 95.17 & 93.60 \\
& Gaming               & 66.85 & 93.42 & 83.22 \\
\cmidrule{2-5}
& \textbf{Overall}     & \textbf{50.15} & \textbf{97.20} & \textbf{92.67} \\
\bottomrule
\end{tabular}}
\caption{Worst, Best, and Average human performance (\%) across tasks and modalities.}
\label{tab:human_performance}
\end{table}

\subsection{C.4 Quantitative GPT Evaluations}
\label{subsec:gpt}

To further assess model performance beyond multiple-choice accuracy, we conduct a quantitative evaluation of the open-ended question responses using GPT-based scoring. Specifically, as discussed in the paper, we prompt GPT-4o to act as an automated grader, evaluating the correctness and relevance of the model's free-form answers based on predefined guidelines consistent with the task objectives in Tab.~\ref{tab:mllm_comparison_gpt} and Tab.~\ref{tab:mllm_comparison_context}.

This automatic assessment provides an additional perspective on the model's generative capabilities, capturing nuances that are not reflected in standard accuracy metrics. The results offer insight into how well the model can produce contextually appropriate and informative responses when not constrained by fixed answer options.

We report the GPT-assigned scores across the full benchmark as well as per agentic task category, enabling a more granular understanding of the model's strengths and weaknesses in open-ended scenarios. These findings complement the multiple-choice evaluations and expose areas where the model demonstrates strong language generation and areas where further improvement is needed. However, we note that these scores may be biased due to limitations in GPT-4o's own reasoning and evaluation capabilities~\cite{li2025preferenceleakagecontaminationproblem}, which can introduce alignment artifacts or systematic preferences.

\begin{table*}[t]
\centering
\renewcommand{\arraystretch}{0.95}
\setlength{\tabcolsep}{6pt}

\begin{tabular}{l ccc ccc}   
\toprule
\multirow{2}{*}{\textbf{Model}} &
  \multicolumn{3}{c}{\textbf{Temporal Consistency}} &
  \multicolumn{3}{c}{\textbf{Correctness}} \\
\cmidrule(lr){2-4}\cmidrule(lr){5-7}
& Video & Audio/Sub & V + A/Sub & Video & Audio/Sub & V + A/Sub \\
\midrule
\textbf{Human}           & 1.85  & 1.69  & 2.83  & 1.93  & 1.46  & 2.65 \\
\midrule
\multicolumn{7}{c}{\textit{Open-source MLLMs}}\\
\midrule
EgoGPT-7B                & 1.3955 & 1.2523 & 1.5634 & 1.2477 & 1.1843 & 1.4930 \\
Ola-7B                   & 0.9178 & 1.2500 & 1.3638 & 0.9038 & 1.2171 & 1.3005 \\
VITA 1.5                 & 0.6643 & 1.2160 & 0.7042 & 0.6561 & 1.0587 & 0.7007 \\
Qwen-2-VL                & 1.4554 & 1.3638 & 1.6772 & 1.3498 & 1.2254 & 1.6092 \\
Qwen-2.5-VL              & 1.6819 & 1.2793 & 1.9730 & 1.4014 & 1.1279 & 1.7477 \\
Qwen-2.5-Omni            & 1.6819 & 1.2793 & 1.9730 & 1.4014 & 1.1279 & 1.7477 \\
InternVL2                & 0.9742 & 1.3169 & 1.1925 & 0.9237 & 0.9143 & 1.1315 \\
LLaVA-OneVision          & 1.5962 & 1.3685 & 1.8052 & 1.4002 & 1.2899 & 1.5763 \\
DeepSeekVL2-small        & 0.9507 & 1.3157 & 1.3286 & 0.9953 & 1.1866 & 1.3650 \\
\midrule
\multicolumn{7}{c}{\textit{Proprietary MLLMs}}\\
\midrule
Gemini 2.0-FL            & 1.6526 & 1.5880 & 2.1784 & 1.4296 & 1.4390 & 1.9014 \\
Gemini 2.5-FL            & 1.7864 & 1.6244 & 2.1678 & 1.5751 & 1.4507 & 1.9167 \\
Claude Sonnet 3.5        & 2.0798 & 1.9155 & 2.6009 & 1.6538 & 1.5857 & 2.2300 \\
GPT-4o                   & 1.5434 & 1.8462 & 2.4390 & 1.4167 & 1.5634 & 2.2207 \\
GPT-4o-mini              & 1.9859 & 1.9272 & 2.3850 & 1.5739 & 1.6185 & 2.0117 \\
Grok 4                   & 2.2782 & 2.1854 & 2.8028 & 1.9507 & 1.7066 & 2.4354 \\
NOVA-Lite                & 0.7453 & 0.9167 & 1.0129 & 0.9812 & 1.1068 & 1.2066 \\
NOVA-Pro                 & 0.9484 & 0.9930 & 1.3333 & 1.1901 & 1.1221 & 1.5387 \\
\bottomrule
\end{tabular}

\caption{GPT-4o evaluation of Temporal Consistency and Correctness scores (higher is better). Columns show performance for \textit{Video}, \textit{Audio/Subtitles}, and their combination.}
\label{tab:mllm_comparison_gpt}
\end{table*}

\begin{table*}[t]
\centering
\renewcommand{\arraystretch}{0.95}
\setlength{\tabcolsep}{6pt}

\begin{tabular}{lcccccc}   
\toprule
\multirow{2}{*}{\textbf{Model}} &
  \multicolumn{3}{c}{\textbf{Context}} &
  \multicolumn{3}{c}{\textbf{Detail Orientation}} \\
\cmidrule(lr){2-4}\cmidrule(lr){5-7}
& Video & Audio/Sub & V\,+A/Sub & Video & Audio/Sub & V\,+A/Sub \\
\midrule
\textbf{Human}            & 2.13  & 1.93  & 3.11  & 1.84  & 2.02  & 2.96 \\
\midrule
\multicolumn{7}{c}{\textit{Open-source MLLMs}} \\
\midrule
EgoGPT-7B                 & 1.6420 & 1.5599 & 1.8732 & 1.2042 & 1.1244 & 1.3509 \\
Ola-7B                    & 1.3146 & 1.6491 & 1.7254 & 1.0047 & 1.2805 & 1.3345 \\
VITA 1.5                  & 0.9906 & 1.4918 & 1.0610 & 0.7465 & 1.2735 & 0.7958 \\
Qwen-2-VL                 & 1.7570 & 1.6549 & 2.0505 & 1.4343 & 1.4178 & 1.7430 \\
Qwen-2.5-VL               & 1.9824 & 1.6197 & 2.2829 & 1.7336 & 1.3885 & 2.0669 \\
InternVL2                 & 1.3462 & 1.3521 & 1.5164 & 1.1303 & 1.0505 & 1.2523 \\
LLaVA-OneVision           & 1.8662 & 1.7136 & 2.0129 & 1.5129 & 1.4284 & 1.7031 \\
DeepSeekVL2-small         & 1.3920 & 1.6984 & 1.7418 & 1.0340 & 1.5998 & 1.3580 \\
\midrule
\multicolumn{7}{c}{\textit{Proprietary MLLMs}} \\
\midrule
Gemini 2.0-FL             & 1.8204 & 1.8439 & 2.1784 & 1.5880 & 1.5293 & 2.0117 \\
Gemini 2.5-FL             & 1.9930 & 1.9120 & 2.3216 & 1.7653 & 1.6678 & 2.0692 \\
Claude Sonnet 3.5         & 2.1725 & 2.0493 & 2.6960 & 2.0129 & 1.9096 & 2.5575 \\
GPT-4o                    & 1.7570 & 2.0634 & 2.6103 & 1.4178 & 1.7946 & 2.3169 \\
GPT-4o-mini               & 2.1444 & 2.1397 & 2.5716 & 1.7277 & 1.8662 & 2.2336 \\
Grok 4                    & 2.5141 & 2.2488 & 2.9636 & 2.4660 & 2.1185 & 2.9566 \\
NOVA-Lite                 & 1.4178 & 1.5516 & 1.6843 & 1.1749 & 1.3451 & 1.3955 \\
NOVA-Pro                  & 1.7312 & 1.5387 & 2.0610 & 1.5282 & 1.3439 & 1.8204 \\
\bottomrule
\end{tabular}

\caption{GPT-4o evaluation of Context score and Detail Orientation score (higher is better). Columns show performance for \textit{Video only}, \textit{Audio/Subtitles only}, and \textit{Video+Audio/Subtitles}.}
\label{tab:mllm_comparison_context}
\end{table*}

\subsection{C.5 Quantitative Evaluations by Categories}
\label{subsec:category}
To provide a comprehensive understanding of model performance across the diverse challenge dimensions of the benchmark, we report quantitative results for all probed models across each defined split. This includes evaluations broken down by agentic task categories, difficulty levels, audio characteristics, topics, and input modalities, as shown in Tab.~\ref{tab:mllm_comparison_category}.

These category-wise results allow us to examine how different models handle specific functional scenarios, content types, and levels of complexity. Through these experiments, we hope to reveal both general patterns and areas of strength or weakness. By comparing performance across these splits, we observe how well models generalize to varying real-world situations and identify which aspects of the benchmark remain most challenging.

Together, these fine-grained evaluations provide a detailed view of model behavior across the full spectrum of tasks and shed light on future work on multimodal understanding and generalization.

\begin{table*}[ht]
\centering
\resizebox{\textwidth}{!}{
\begin{tabular}{llcccccc}
\toprule
\textbf{} & \textbf{} 
& \textbf{GPT-4o} 
& \textbf{Gemini 2.5 FL} 
& \textbf{Qwen 2.5 Omni} 
& \textbf{Nova Lite} 
& \textbf{Nova Pro} 
& \textbf{Ola-7B} \\
\midrule

\multirow{3}{*}{\rotatebox{90}{\textit{Difficulty}}}
& \textit{Easy}    & 66.7 & 59.8 & 52.1 & 55.0 & 62.5 & 57.4 \\
& \textit{Medium}  & 67.4 & 54.9 & 48.8 & 52.4 & 54.9 & 54.1 \\
& \textit{Hard}    & 49.7 & 43.0 & 45.2 & 38.4 & 43.0 & 41.1 \\

\midrule
\multirow{7}{*}{\rotatebox{90}{\textit{Agentic Tasks}}}
& Social                 & 62.6 & 55.6 & 47.5 & 48.5 & 50.5 & 52.5 \\
& Sentiment              & 56.8 & 54.3 & 58.0 & 42.0 & 55.6 & 55.6 \\
& Egocentric Agent       & 59.3 & 39.0 & 42.4 & 44.1 & 54.2 & 57.6 \\
& Information Querying   & 68.3 & 60.0 & 45.3 & 52.1 & 54.7 & 49.1 \\
& Sports                 & 60.5 & 47.1 & 49.6 & 50.4 & 58.0 & 52.9 \\
& Gaming                 & 61.2 & 49.5 & 41.7 & 39.8 & 47.6 & 46.6 \\
& Shopping               & 68.3 & 61.9 & 64.3 & 69.0 & 67.5 & 63.5 \\

\midrule
\multirow{5}{*}{\rotatebox{90}{\textit{Audio Type}}}
& Natural Sound    & 52.9 & 51.5 & 52.9 & 35.3 & 45.6 & 48.5 \\
& Speech          & 66.4 & 56.7 & 47.8 & 53.8 & 57.6 & 52.0 \\
& Music           & 61.5 & 50.0 & 53.8 & 42.3 & 57.7 & 38.5 \\
& Artificial Sound & 64.0 & 52.0 & 48.0 & 48.0 & 50.0 & 58.0 \\
& Mixed Sounds    & 63.0 & 53.1 & 51.5 & 51.5 & 56.1 & 56.5 \\

\midrule
\multirow{13}{*}{\rotatebox{90}{\textit{Topics}}}
& Humanities and Society     & 71.4 & 64.7 & 56.4 & 54.1 & 63.9 & 54.1 \\
& Geography and Travel       & 56.2 & 50.0 & 37.5 & 56.2 & 31.2 & 56.2 \\
& Technology and Gaming      & 65.6 & 55.0 & 42.0 & 43.5 & 48.9 & 48.9 \\
& Science and Knowledge      & 60.7 & 50.0 & 50.0 & 46.4 & 35.7 & 60.7 \\
& Movies, TV and Animations  & 53.7 & 37.3 & 34.3 & 43.3 & 52.2 & 38.8 \\
& Arts and Performance       & 51.7 & 44.8 & 51.7 & 44.8 & 44.8 & 48.3 \\
& Pets and Animals           & 57.8 & 66.7 & 55.6 & 40.0 & 53.3 & 53.3 \\
& Business and Commerce      & 69.8 & 66.3 & 68.6 & 73.3 & 69.8 & 66.3 \\
& Life and Practical Skills  & 72.1 & 60.6 & 51.9 & 60.6 & 61.5 & 61.5 \\
& Sports and Adventure       & 61.1 & 49.2 & 48.4 & 46.0 & 58.7 & 50.0 \\
& Social Trends and Reactions & 53.7 & 48.1 & 44.4 & 46.3 & 40.7 & 50.0 \\
& Vehicles and Transportation & 66.7 & 33.3 & 55.6 & 44.4 & 77.8 & 66.7 \\
& Low Quality and Extended Content & 58.3 & 29.2 & 25.0 & 41.7 & 50.0 & 37.5 \\

\midrule
\multirow{8}{*}{\rotatebox{90}{\textit{Multimodal Abilities}}}
& Visual Reasoning     & 58.3 & 54.4 & 45.6 & 50.0 & 51.8 & 52.6 \\
& Emotional Inference  & 53.5 & 65.1 & 51.2 & 39.5 & 44.2 & 44.2 \\
& Situational Reasoning & 70.0 & 65.0 & 35.0 & 55.0 & 65.0 & 70.0 \\
& Causal Reasoning    & 67.2 & 49.3 & 44.8 & 49.3 & 55.2 & 47.8 \\
& Factual Recall      & 67.3 & 55.4 & 52.9 & 53.8 & 58.6 & 54.2 \\
& Spatial Reasoning   & 55.2 & 41.4 & 48.3 & 34.5 & 55.2 & 37.9 \\
& Relationship        & 68.8 & 50.0 & 50.0 & 56.2 & 62.5 & 75.0 \\
& Temporal Distance   & 75.0 & 50.0 & 50.0 & 41.7 & 50.0 & 58.3 \\

\midrule
& Overall & 64.0 & 54.7 & 49.5 & 50.9 & 55.8 & 53.1 \\

\bottomrule
\end{tabular}
}
\caption{Performance comparison across different models and categories with \textit{Video + Audio/Subtitle} modalities.}
\label{tab:mllm_comparison_category}
\end{table*}

\subsection{C.6 Common Error Modes}
\label{subsec:error}
\begin{figure*}[t]
    \centering
    \includegraphics[width=0.8\textwidth]{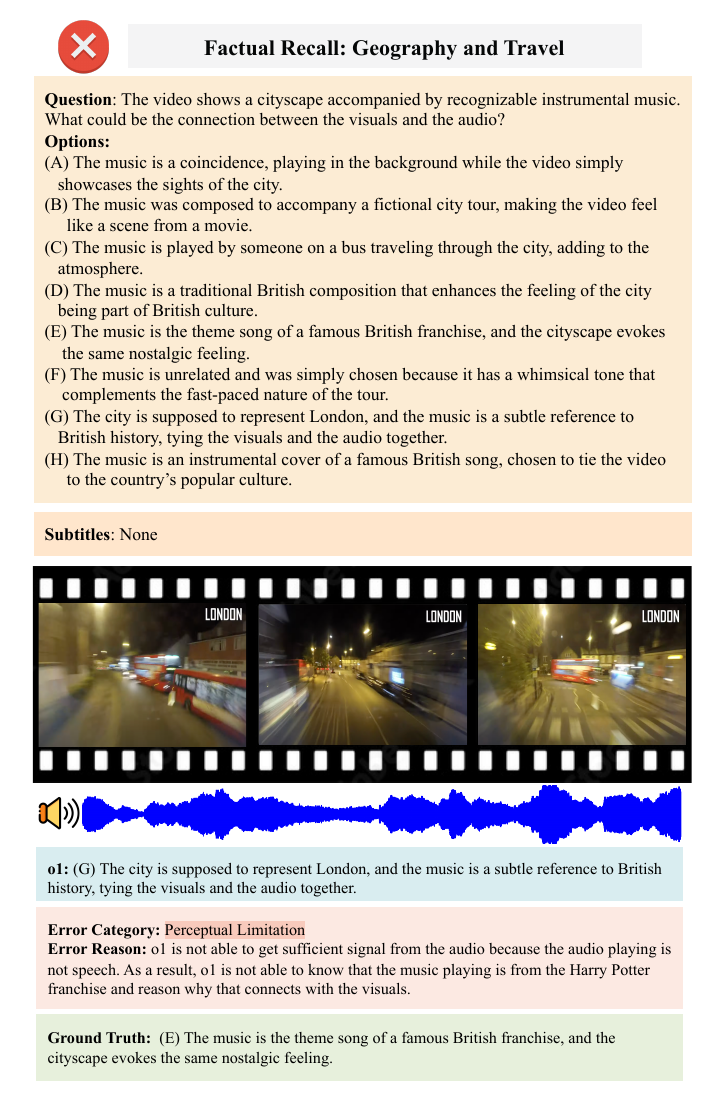}
  \caption{Error analysis showing that \textit{o1} fails to correctly answer the question when the audio cannot be transcribed into text-based subtitles, leading to an incorrect connection between the cityscape video and the instrumental music.}
  \label{fig:wrong_answer1}
\end{figure*}

\begin{figure*}[t]
    \centering
    \includegraphics[width=0.8\textwidth]{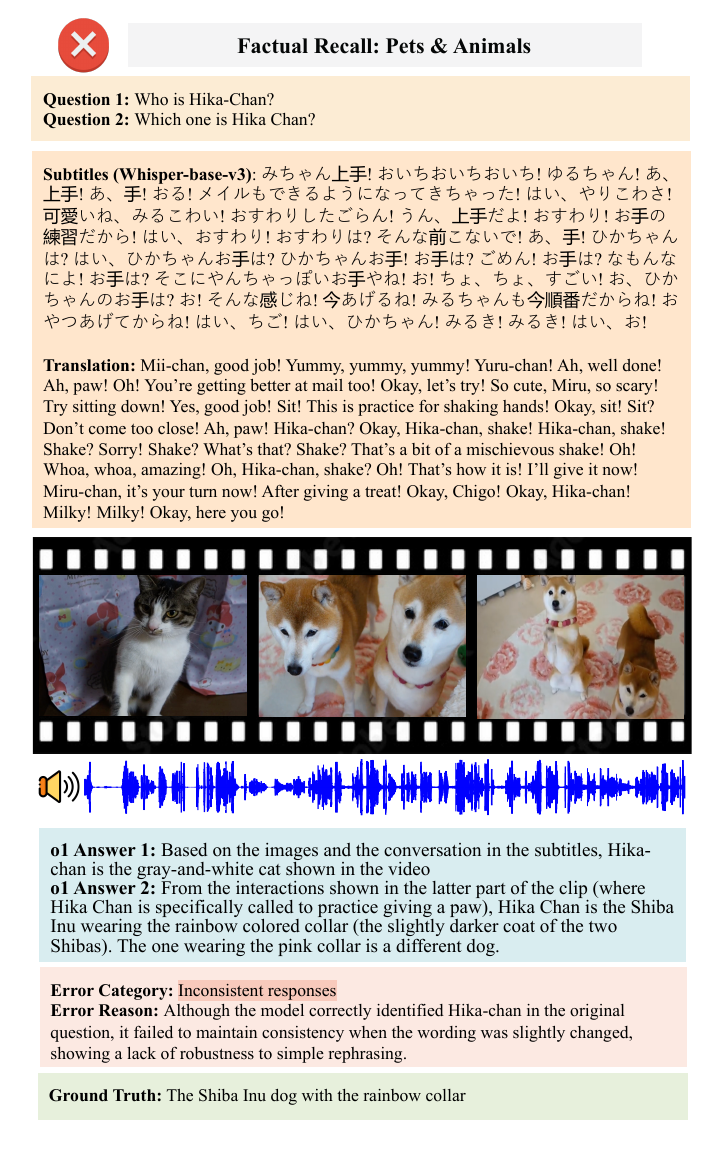}
  \caption{Error analysis showing inconsistent responses from \textit{o1}, where it failed to accurately identify Hika-chan, when the question was rephrased slightly.}
  \label{fig:wrong_answer2}
\end{figure*}

\begin{figure*}[t]
    \centering
    \includegraphics[width=0.8\textwidth]{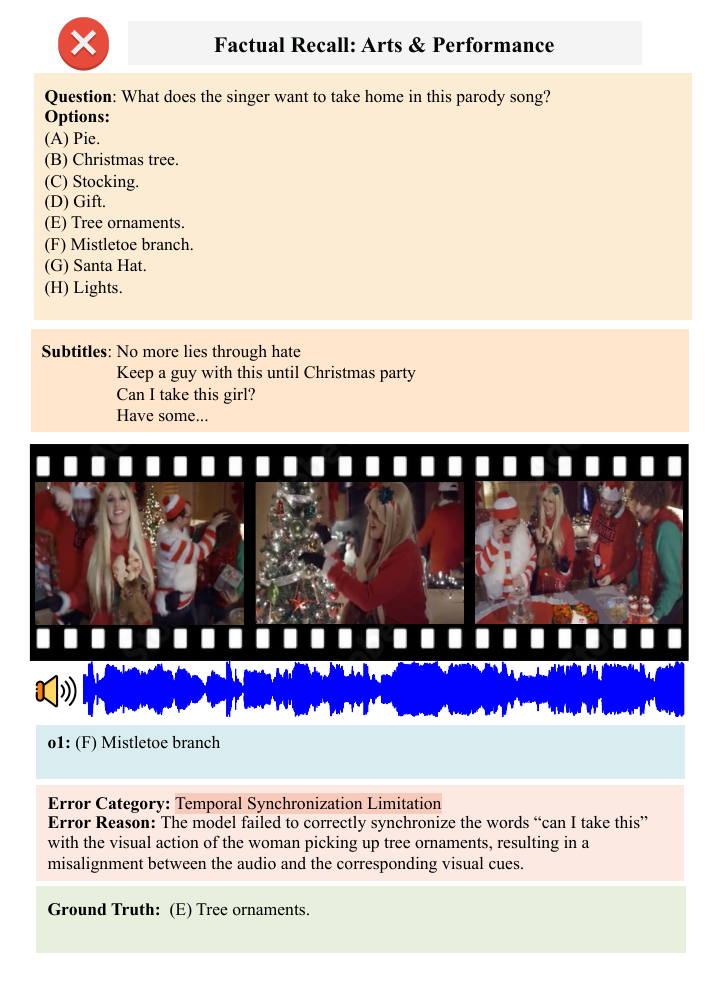}
  \caption{Error analysis highlighting \textit{o1}’s failure in correctly synchronizing the audio with visual cues, leading to misinterpretation of the woman picking up tree ornaments.}
  \label{fig:wrong_answer3}
\end{figure*}

\begin{figure*}[t]
    \centering
    \includegraphics[width=0.8\textwidth]{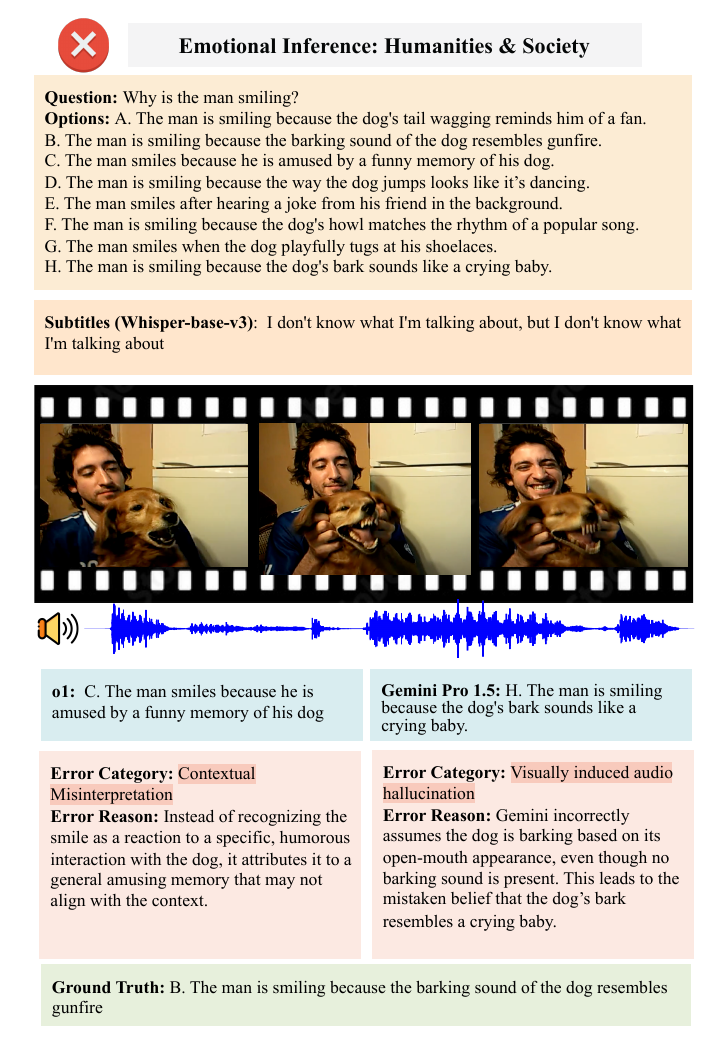}
  \caption{Error analysis highlighting \textit{o1}’s contextual misinterpretation, attributing the man’s smile to a general amusing memory rather than the dog’s barking resembling gunfire. Additionally, \textit{Gemini Pro 1.5} exhibits visually induced audio hallucination, mistaking the dog’s open-mouth appearance for barking and associating it with a crying baby.}
  \label{fig:wrong_answer4}
\end{figure*}

\begin{figure*}[t]
    \centering
    \includegraphics[width=0.8\textwidth]{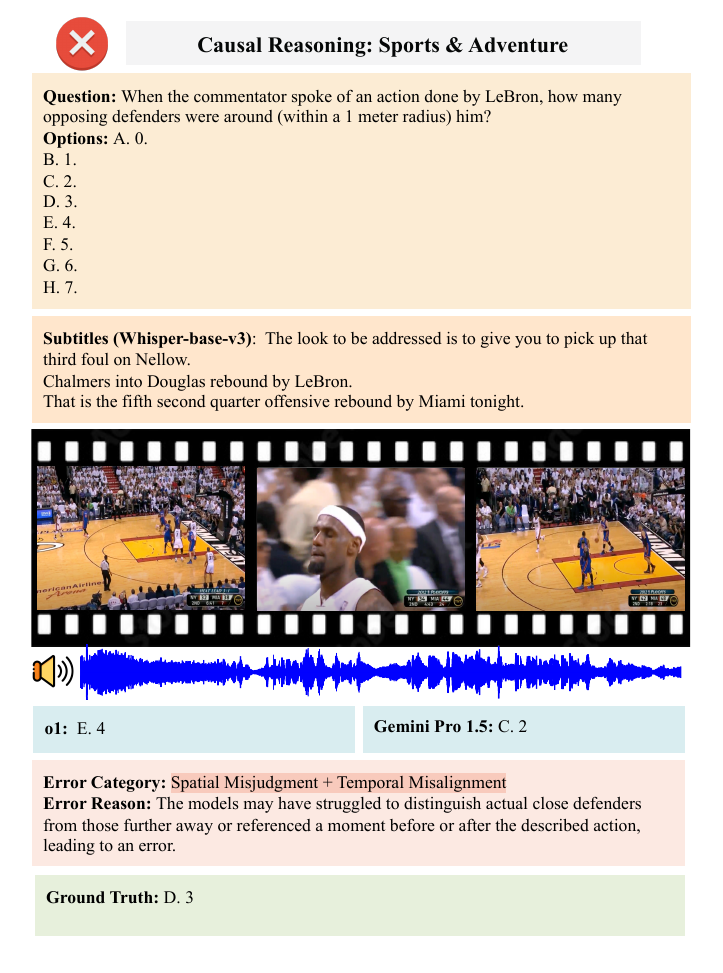}
  \caption{Error analysis showing \textit{o1} and \textit{Gemini Pro 1.5} both struggling with spatial misjudgment and temporal misalignment, leading to errors in estimating the number of defenders due to difficulties in distinguishing actual close defenders and potential misalignment with the described action.}
  \label{fig:wrong_answer5}
\end{figure*}

\begin{figure*}[t]
    \centering
    \includegraphics[width=0.8\textwidth]{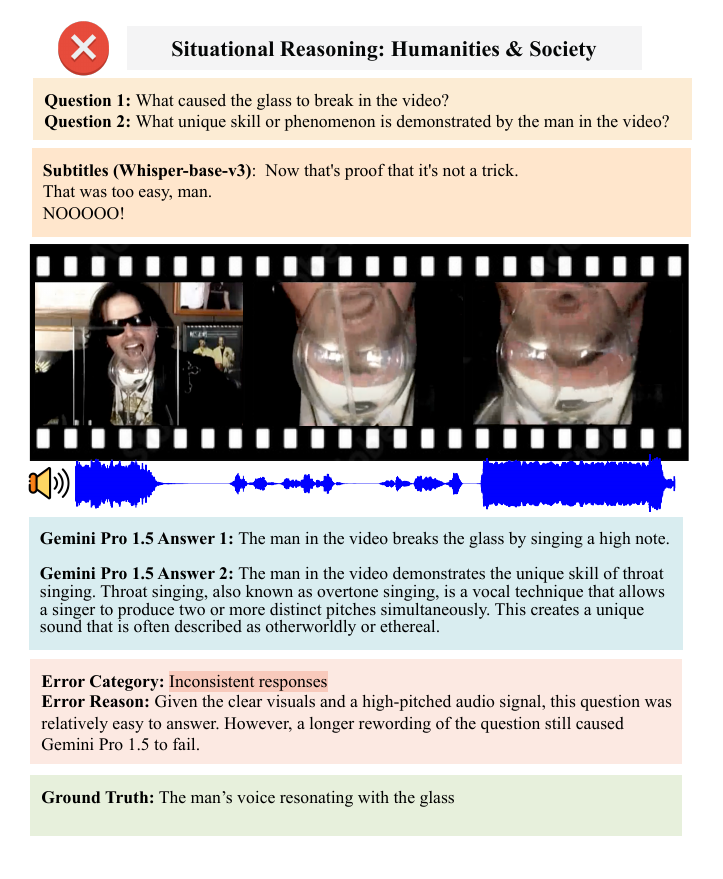}
  \caption{Error analysis showing \textit{Gemini Pro 1.5} providing inconsistent responses despite clear visuals and audio cues, failing to accurately identify the cause of the glass breaking due to rephrased question structure.}
  \label{fig:wrong_answer6}
\end{figure*}

To better understand the limitations of the state of the art models, we analyze common error patterns observed across the benchmark. We find that model failures often arise from modality-specific weaknesses, such as misinterpretation of complex visual cues, inability to process nuanced audio signals, or challenges in aligning multimodal information with nuanced context. For illustration, we highlight representative failure cases from OpenAI o1 and Gemini 1.5 Pro, chosen because they are among the strongest available models; however, the same error patterns recur across the models we evaluated. Common failure modes are showcased in Fig.~\ref{fig:wrong_answer1}, ~\ref{fig:wrong_answer2}, ~\ref{fig:wrong_answer3}, ~\ref{fig:wrong_answer4}, ~\ref{fig:wrong_answer5}, ~\ref{fig:wrong_answer6}.

\section{D. Questionnaire Survey}
\label{sec:survey}

To establish human performance baselines for comparison with the probed models, we conducted a questionnaire study approved by our institutional IRB. We recruited participants with proficiency in at least one language present in the dataset and prior familiarity with video content to ensure they could reasonably complete the tasks. Participants answered questions presented in the same format as those used for model evaluation.

The survey included a total of 261 questions curated from the dataset, covering diverse subcategories within each defined split. For each participant, the questionnaire consisted of either one multiple-choice question (MCQ) with eight answer options, or an open-ended question from a different video. To minimize content overlap and reduce potential bias, no video was reused between questions within a single questionnaire. This approach ensured broad coverage of the dataset's taxonomy while avoiding contamination across tasks. We recruited 382 participants through Amazon Mechanical Turk service with approval rate $>$96\% to answer the questions with A, S, V, A+V, S+V as inputs with media from MAVERIX and MAVERIX-Long. Both MCQ responses and open-ended responses are recorded. 

For evaluation, MCQ responses were scored against the ground truth to calculate accuracy, while open-ended answers were assessed using the same GPT-4o evaluation pipeline applied to model outputs, ensuring consistency across human and model scoring.

\begin{figure*}[t]
  \includegraphics[width=\textwidth]{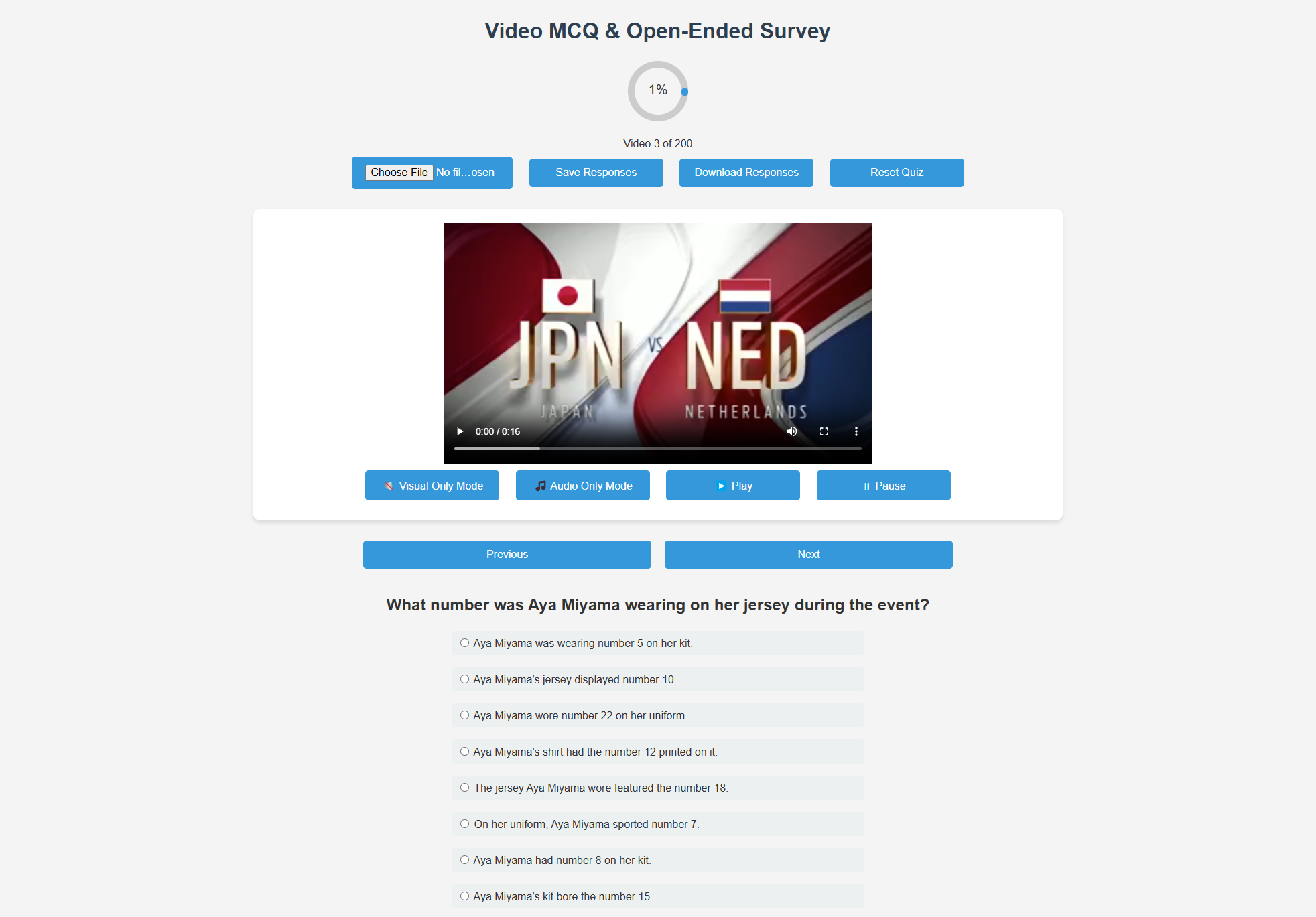}
  \caption{Screenshot of the questionnaire under the audio-plus-visual condition, showcasing the interface used to assess human performance across multiple-choice questions. Results offer a baseline for comparison with multimodal model performance across diverse tasks and difficulty levels.}
  \label{fig:sup_both}
\end{figure*}

\begin{figure*}[t]
  \includegraphics[width=\textwidth, trim=0 320 0 0, clip]{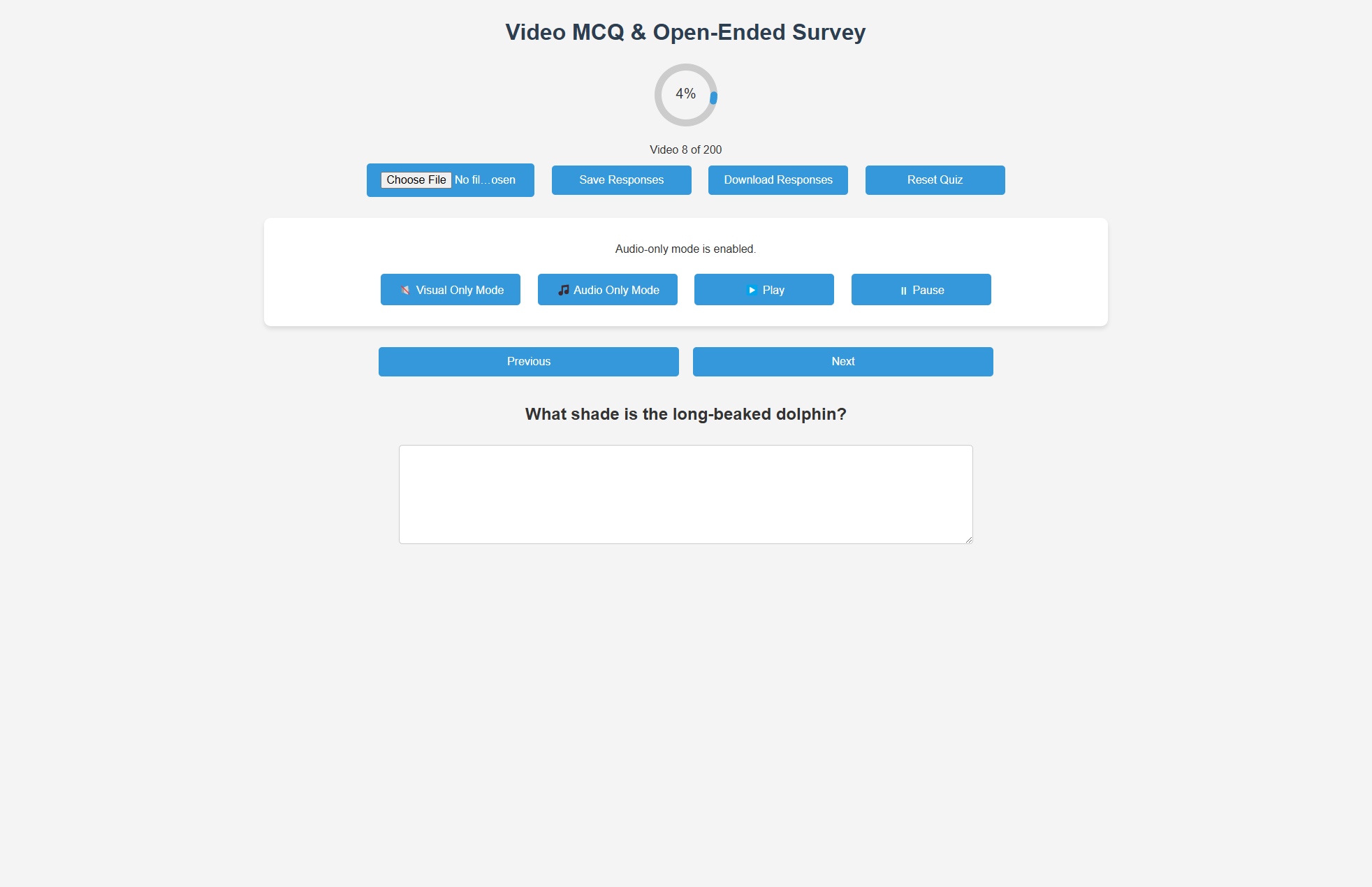}
  \caption{Questionnaire screenshot showing open-ended questions under the audio-only condition. Responses were evaluated using the GPT-4o grading pipeline, capturing the quality of language generation based solely on audio context.}
    \label{fig:sup_audio}

\end{figure*}

\begin{figure*}[t]
  \includegraphics[width=\textwidth]{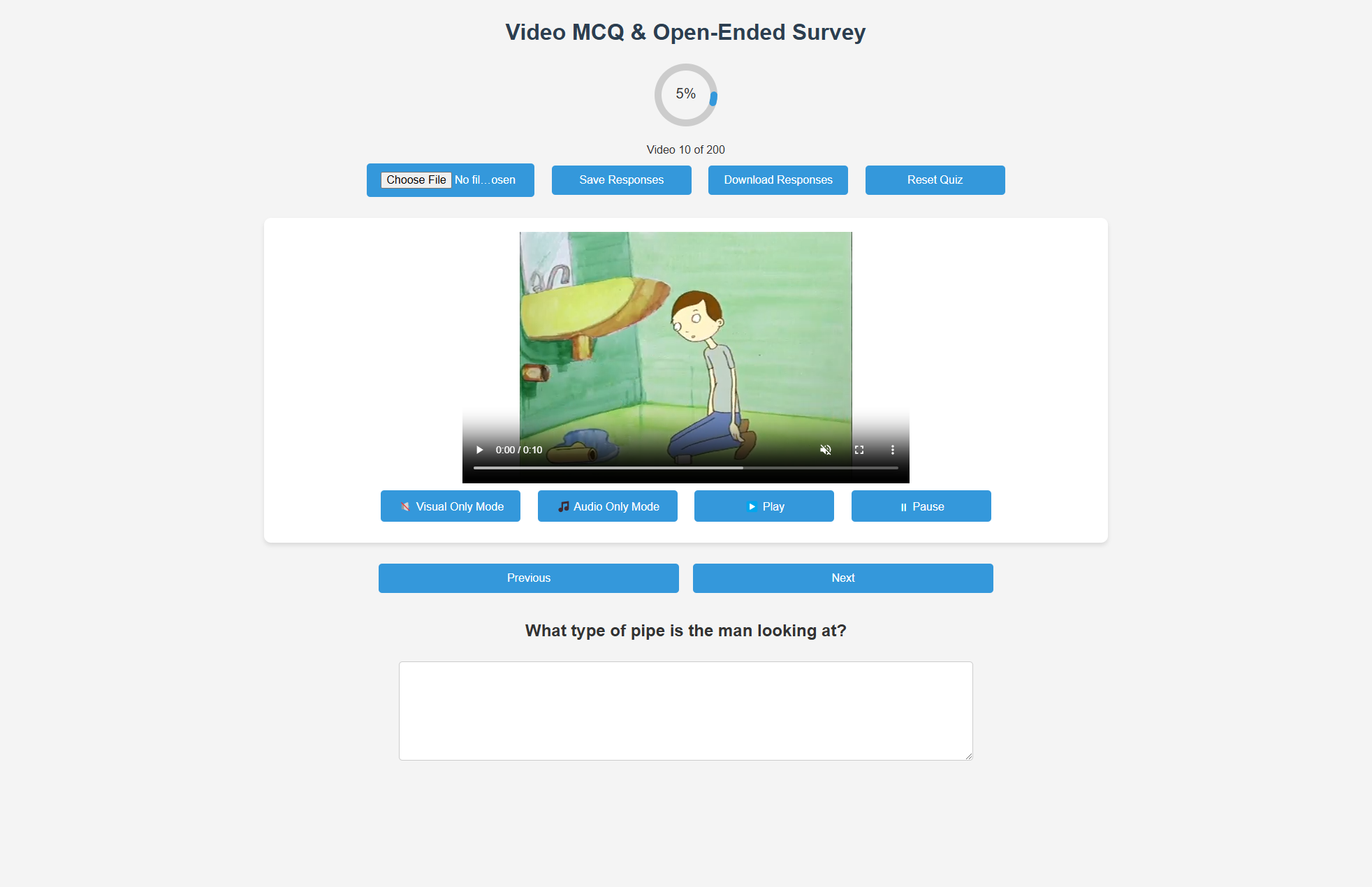}
  \caption{Questionnaire screenshot showing open-ended questions under the visual-only condition. GPT-4o evaluation scores assess human understanding based solely on visual information, without the aid of audio context.}
  \label{fig:sup_visual}
\end{figure*}

We tested human performance across three different conditions: audio only, visual only, and audio plus video. Each condition was toggle-enabled within the survey toolkit, allowing participants to be assigned to a specific modality. The survey interfaces for the conditions are shown in Fig. \ref{fig:sup_audio} for audio only, Fig. \ref{fig:sup_visual} for visual only, Fig. \ref{fig:sup_both} for both modalities. To prevent cross-condition contamination and ensure focused evaluation, each participant completed the entire study under only one selected modality.

This human study provides us with valuable reference points for interpreting model results and understanding how humans perform under comparable multimodal constraints.


\section{E. Dataset Distribution Continued}
\label{sec:distribution}
\begin{figure*}[t]
  \includegraphics[width=0.95\textwidth]{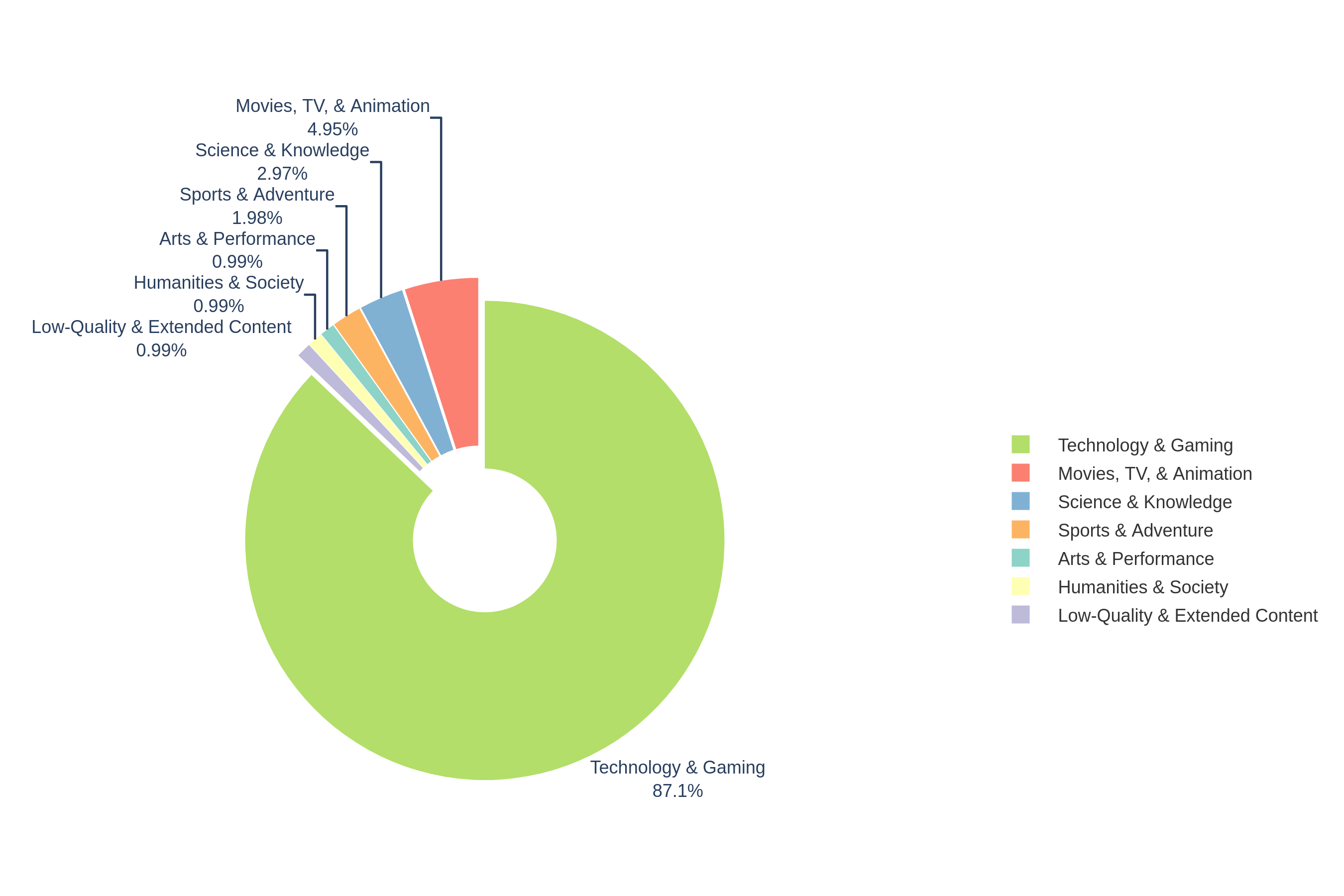}
  \caption{Taxonomy composition in Gaming.}
  \label{fig:sup_gaming}
\end{figure*}

\begin{figure*}[t]
  \includegraphics[width=0.9\textwidth, trim=0 80 0 100, clip]{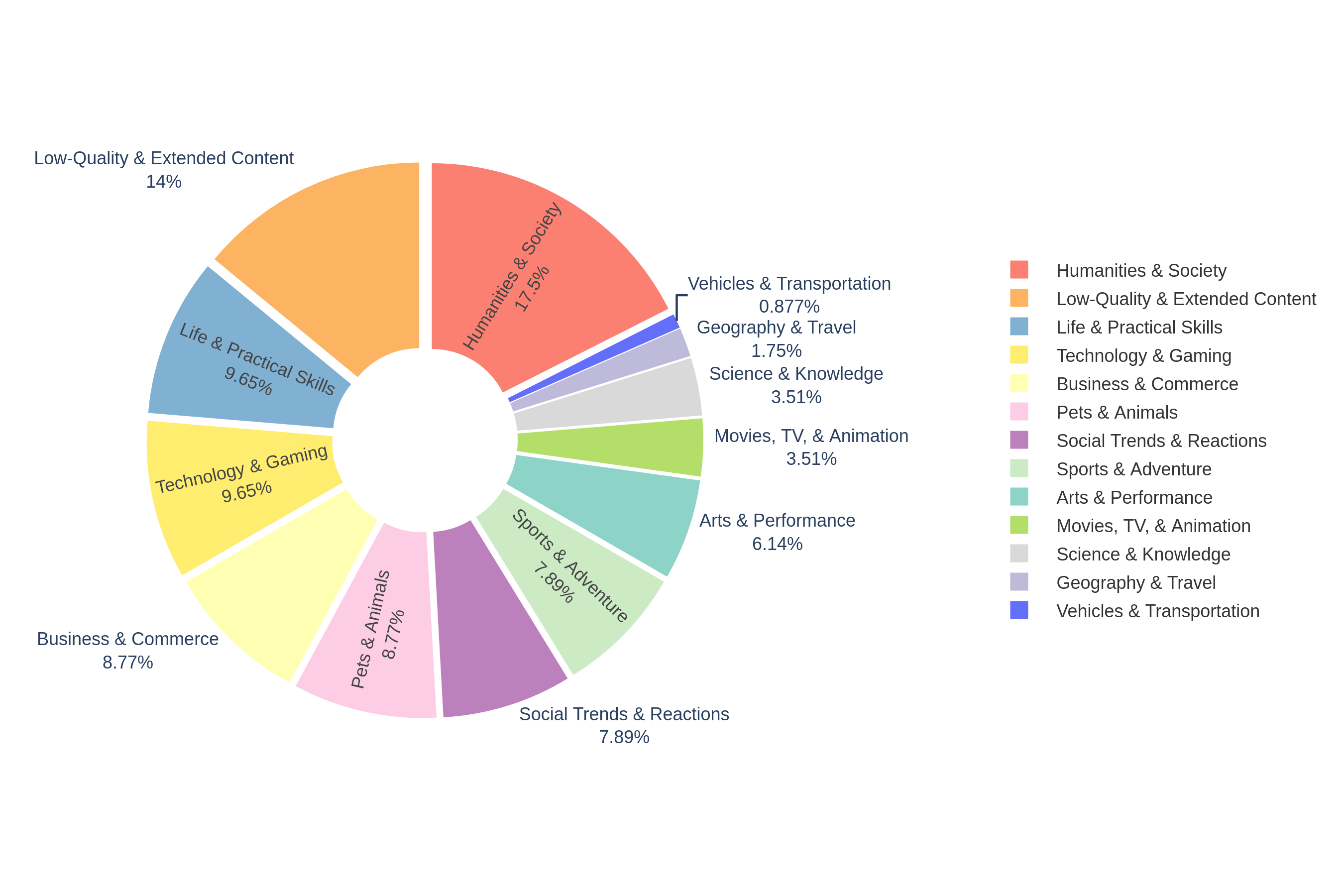}
  \caption{Taxonomy composition in Egocentric Agent.}
  \label{fig:sup_egocentric}
\end{figure*}

\begin{figure*}[t]
  \includegraphics[width=0.8\textwidth, trim=0 80 0 100, clip]{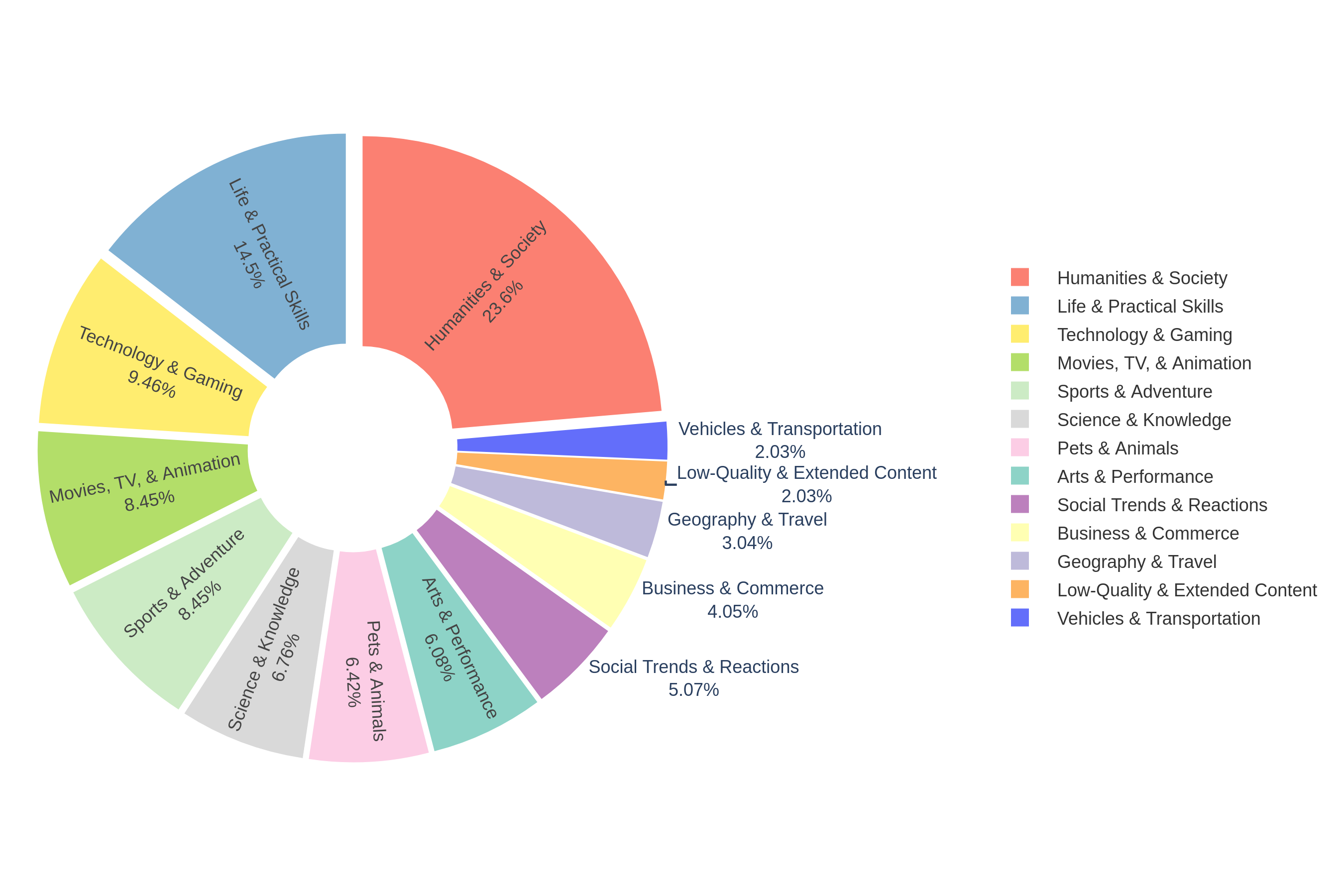}
  \caption{Taxonomy composition in Information Querying.}
  \label{fig:sup_info}
\end{figure*}

\begin{figure*}[t]
  \includegraphics[width=\textwidth, trim=0 80 0 100, clip]{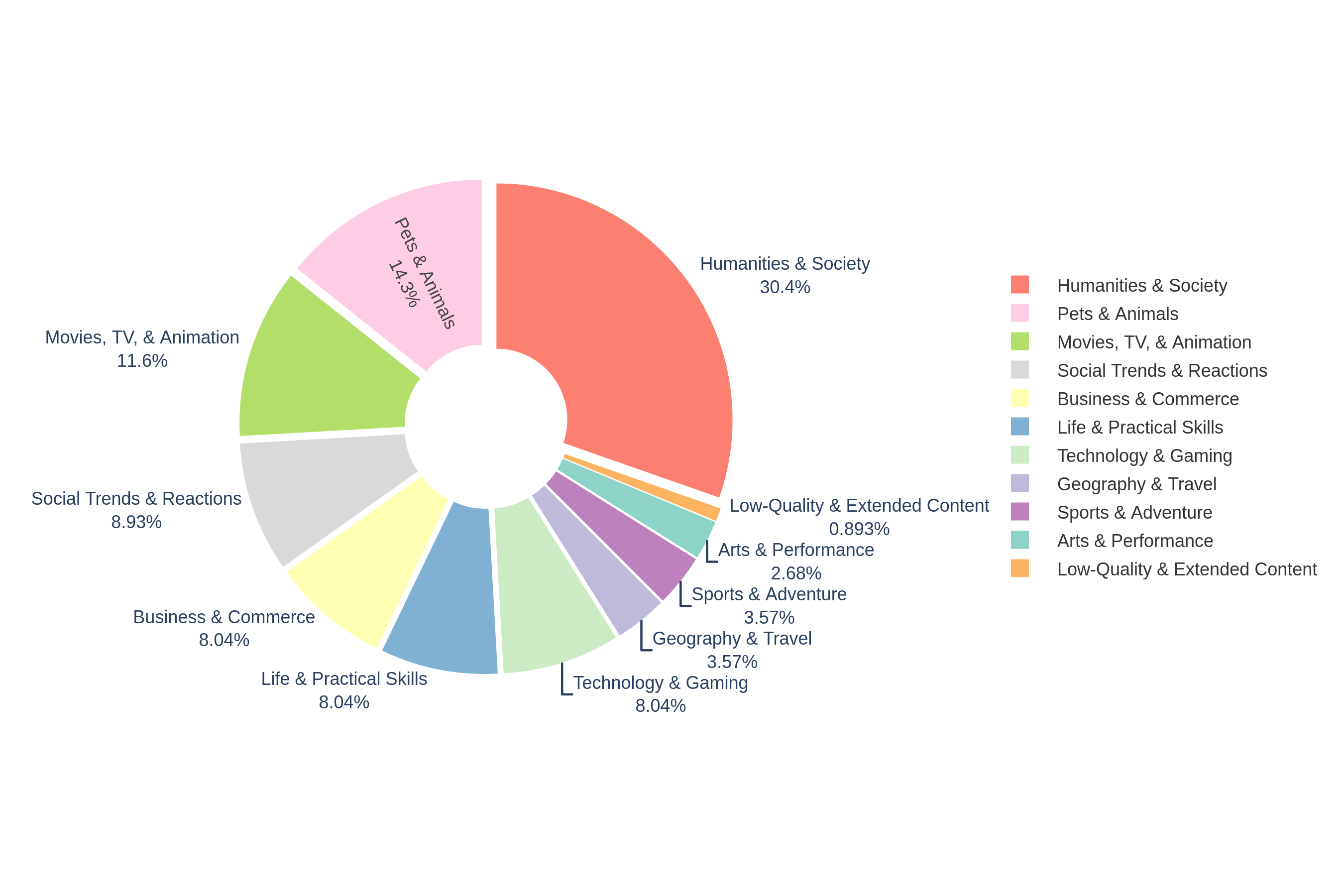}
  \caption{Taxonomy composition in Sentiments.}
  \label{fig:sup_sentiments}
\end{figure*}

\begin{figure*}[t]
  \includegraphics[width=0.9\textwidth, trim=0 80 0 100, clip]{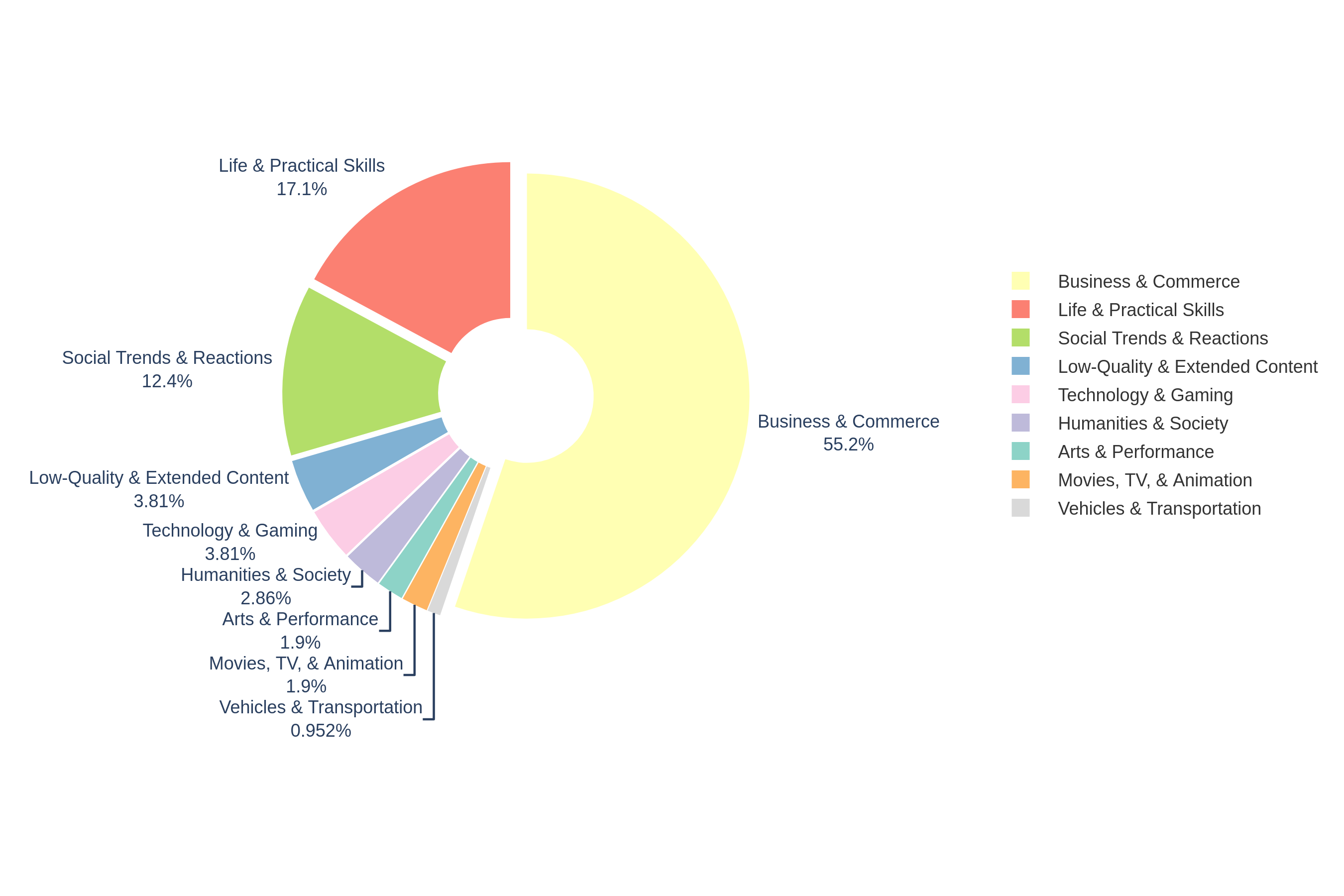}
  \caption{Taxonomy composition in Shopping.}
  \label{fig:sup_shopping}
\end{figure*}

\begin{figure*}[t]
  \includegraphics[width=0.9\textwidth, trim=0 80 0 100, clip]{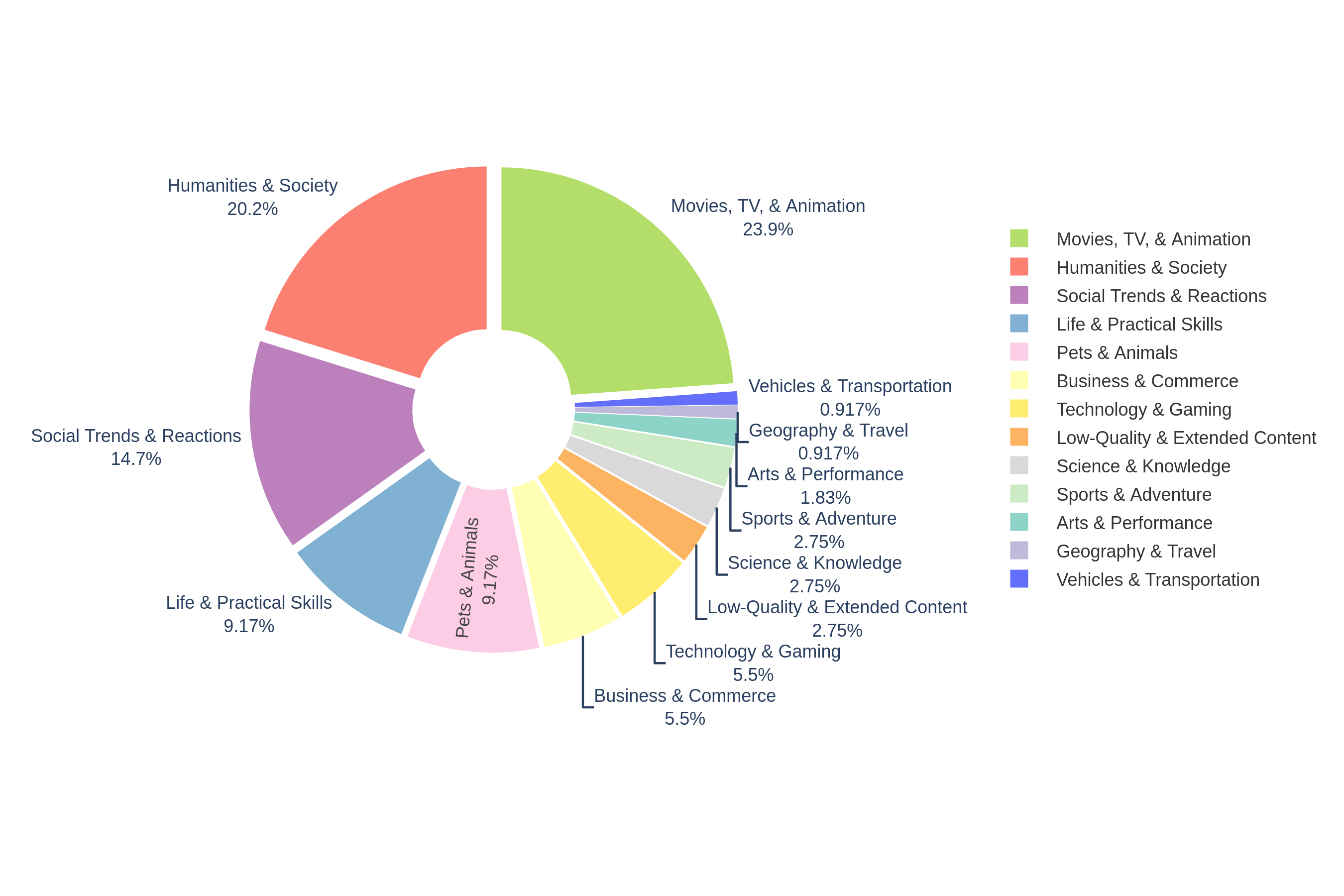}
  \caption{Taxonomy composition in Social.}
  \label{fig:sup_social}
\end{figure*}

\begin{figure*}[t]
  \includegraphics[width=0.9\textwidth, trim=0 80 0 100, clip]{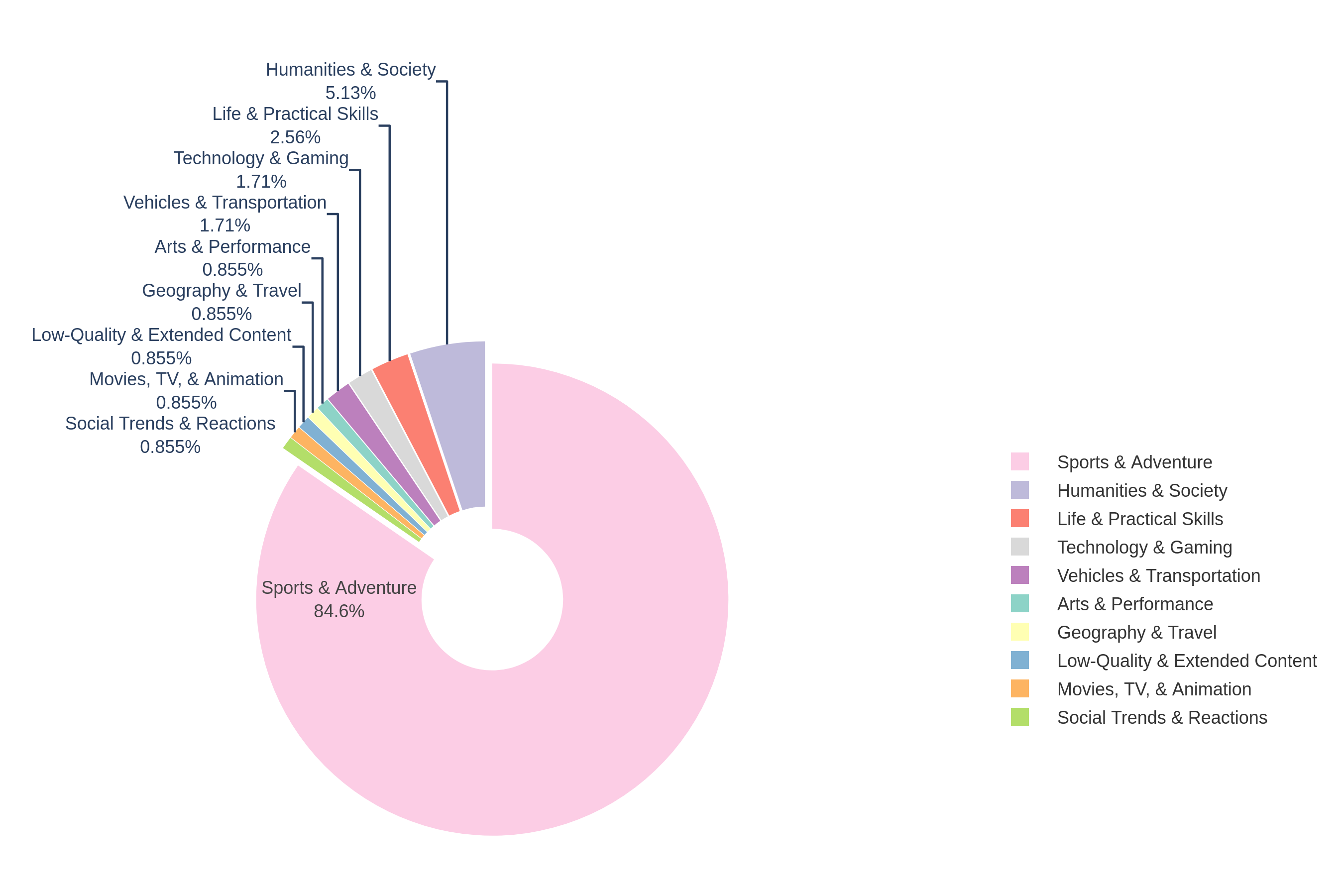}
  \caption{Taxonomy composition in Sports.}
  \label{fig:sup_sports}
\end{figure*}

In this section, we provide a detailed breakdown of the taxonomy distributions within each defined agentic task category. For each task type, we analyze the composition of its associated taxonomies to illustrate the diversity and balance of the dataset across different functional scenarios. The corresponding distributions are visualized in Fig. \ref{fig:sup_gaming} for Gaming, Fig. \ref{fig:sup_egocentric} for Egocentric Agent, Fig. \ref{fig:sup_info} for Information Querying, Fig. \ref{fig:sup_sentiments} for Sentiments, Fig. \ref{fig:sup_shopping} for Shopping, Fig. \ref{fig:sup_social} for Social, and Fig. \ref{fig:sup_sports} for Sports. These visualizations offer insight into the internal structure of each agentic task and highlight the variety of situations represented within the dataset.

\clearpage


\end{document}